\newcommand{\cL}{{\cal L}}
\newcommand{\cA}{{\cal A}}
\newcommand{\bq}{{\bf q}}
\newcommand{\bal}{{\bf \alpha}}

\documentclass[12pt]{iopart}
\usepackage{graphicx,cite}
\usepackage{amssymb}

\begin{document}

\title[Regular spatial structures in arrays of BEC]
      {Regular spatial structures in arrays of Bose-Einstein
      condensates induced by modulational instability}

\author{B B Baizakov\dag\footnote[3]
       {Permanent address: Physical-Technical Institute,
       2-b Mavlyanov str., 700084 Tashkent Uzbekistan},
       \ V V Konotop\ddag\ and M Salerno\dag}

\address{\dag\ Dipartimento di Fisica "E.R. Caianiello"
         and Istituto Nazionale di Fisica della Materia (INFM),
         Universit\'a di Salerno, I-84081 Baronissi (SA), Italy}

\address{\ddag\ Departmento de F\'{\i}sica and Centro de F\'{\i}sica da
         Materia Condensada, Universidade de Lisboa,
         Complexo Interdisciplinar, Av. Prof. Gama Pinto 2, Lisboa 1649-003,
         Portugal}

\date{\today}

\eads{\mailto{baizakov@sa.infn.it},
      \mailto{konotop@cii.fc.ul.pt},
      \mailto{salerno@sa.infn.it}}


\begin{abstract}
We show that the phenomenon of modulational instability in arrays
of Bose-Einstein condensates confined to optical lattices gives
rise to coherent spatial structures of localized excitations.
These excitations represent thin disks in 1D, narrow tubes in 2D,
and small hollows in 3D arrays, filled in with condensed atoms of
much greater density compared to surrounding array sites. Aspects
of the developed pattern depend on the initial distribution
function of the condensate over the optical lattice, corresponding
to particular points of the Brillouin zone. The long-time behavior
of the spatial structures emerging due to modulational instability
is characterized by the periodic recurrence to the initial
low-density state in a finite optical lattice. We propose a simple
way to retain the localized spatial structures with high atomic
concentration, which may be of interest for applications.
Theoretical model, based on the multiple scale expansion,
describes the basic features of the phenomenon. Results of
numerical simulations confirm the analytical predictions.
\end{abstract}

\submitto{\jpb}

\section{Introduction}
\label{introduction}

Optical lattices formed by laser waves are the media where
Bose-Einstein condensates (BEC) exhibit remarkable properties.
Over the last few years there has been a significant progress in
manipulation with BEC confined to optical lattices which resulted
in observation of diverse new phenomena, such as coherent emission
of Bose-condensed atoms \cite{anderson}, Bloch oscillations and
Landau-Zener tunneling \cite{morsch}, atomic Josephson effect
\cite{cataliotti}, Mott insulator - superfluid transition
\cite{grainer1}. Realization of BEC arrays in 2D and 3D optical
lattices \cite{grainer1,grainer2} has opened new perspectives for
investigation of fundamental properties of quantum gases in lower
dimensions. Theoretical studies on the dynamics of BEC in a
periodic potential critically rely on the concepts of band
structure and Bloch states, acquired from the solid state physics.
Recent developments in the field show that the above concepts,
originally constructed for linear periodic systems, also play the
crucial role in the physics of nonlinear periodic systems, such as
BEC arrays in optical lattices \cite{wu,steel,ks}. A good
correlation between predictions of the band theory for the BEC
dynamics in a periodic potential and experimental results with
static, moving, and accelerating 1D optical lattices was
demonstrated in \cite{denschlag}. Wide range of spatiotemporal
behaviour of a BEC in 1D linear- and circular-chain optical
lattices in the tight-binding limit, when the dynamics is
described by the discrete nonlinear Schr\"odinger equation, was
numerically investigated in Ref.\cite{tsukada}. Transmission of
matter wave pulses incident in a 1D optical lattice, including
their collision dynamics, was theoretically considered in
\cite{carusotto}.

A subject, which is interesting both from the viewpoints of the
theory and applications of BEC, is the origin and dynamics of
spatially localized nonlinear excitations in the condensate
confined to a periodic potential. A stimulating discovery was the
proof of the existence of bright solitons in the effectively 1D
BEC arrays with repulsive interaction between atoms
\cite{ks,alfimov,potting}. In view of the fact that a continuous
BEC with repulsive interatomic forces does not support spatially
localized humps of atomic concentration, BEC arrays in optical
lattices are considered to be the most inviting media for the
creation and manipulation of soliton-like structures with
ultracold atoms. The physical mechanism by which this possibility
arises is similar to that of electrons in a periodic potential, in
specific cases acquiring negative effective mass. The presence of
the optical lattice can invert the sign of the dispersive term,
which then balances the action of the nonlinearity. Therefore,
bright solitons in BEC arrays with repulsive interaction between
atoms are possible in the presence of the periodic potential of
the optical lattice. It is worth to mention here that although a
continuous BEC with attractive interaction between atoms can bear
bright solitons, its other property leading to a collapse of the
condensate at some critical atomic concentration (for review see
e.g. \cite{dalfovo}), makes it less appealing for the above
purpose. In recent papers \cite{khaykovich,strecker}, reporting on
the first experimental observation of matter-wave bright solitons
in a continuous BEC with attractive interatomic forces ($^7$Li),
the macroscopic quantum bound state of Bose-condensed atoms
(bright soliton) was shown to exist in a narrow window of
atomic numbers (around ${\mathcal N}  \sim 5000$). Beyond that window atomic
wavepackets undergo collapse or explosion. Moreover, the bright
soliton with attractive forces between atoms subject to expulsive
potential, as applied in \cite{khaykovich,strecker}, appears to be
of limited lifetime due to the effect of quantum tunneling
\cite{carr} (termed by authors as {\it quantum evaporation}),
which leads to eventual explosion of the soliton. Therefore,
matter-wave bright solitons composed of repulsive atoms in optical
lattices, which are free of the above constraints, seem to be
advantageous for applications.

Out of existing studies on localized excitations in BEC arrays, little
attention has been devoted to methods of creation of such structures, so far.
It has recently been suggested to employ the modulational instability,
which constitutes one of the most important phenomena associated with the
evolution of nonlinear waves, to create bright BEC solitons in a 1D optical
lattice \cite{ks}. A variety of localized solutions are found to the
one-dimensional nonlinear Schr\"odinger equation with a periodic potential,
some of which are spatially and temporally stable \cite{alfimov}.
Interesting consequence of the modulational instability in
a continuous BEC was reported in \cite{robins}. The authors have
shown that the modulational instability leads to fragmentation of the
ferromagnetic phase in a spinor Bose-condensate. Another manifestation
of the modulational instability as leading to dynamical superfluid-insulator
transition in a BEC confined to an optical lattice and magnetic potential
has been studied in \cite{smerzi2002}.

It is appropriate to mention, that the phenomenon of modulational instability
is well studied in different areas of nonlinear physics, since initiated in
the 1960s, by predictions in hydrodynamics \cite{benjamin},
plasmas \cite{taniuti,askaryan}, and nonlinear optics
\cite{ostrovskii,karpman}.
For later reviews on modulational instability in Hamiltonian systems the
reader is addressed to references \cite{kuznetsov,berge,akhmediev}.
In view of the existence of many features of ultracold atomic gases similar
to those observed in the above mentioned systems, there is a solid ground to
expect rich dynamics induced by the modulational instability in such a
nonlinear system as Bose-Einstein condensates.

In the present paper we study the dynamical processes in BEC
arrays confined to one-, two-, and three-dimensional optical
lattices, which are due to the modulational instability.
Particularly we focus on the coherent spatial structures in 2D and
3D BEC arrays, which originate from the modulational instability.

The paper is organized as follows: Section \ref{model} contains the
derivation of main equations, as well as brief exposition of the
modulational instability. In section \ref{bands} we present the
energy band structure for a BEC distributed over the periodic
potential. In section \ref{numerics} we analyze the features of
spatial structures in arrays of BEC, originated form the
modulational instability of the initial waveforms, corresponding
to different points of the Brillouin zone. Section
\ref{conclusions} summarizes the results of this study.

\section{The multiscale analysis and modulational instability}
\label{model}

To develop the model we consider a dimensionless
3D Gross-Pitaevskii (GP) equation
\begin{equation} \label{NLS_per}
  i\frac{\partial \psi}{\partial t} = -\Delta \psi+
  V({\bf r}) \psi + \chi |\psi|^2 \psi,
\end{equation}
where ${\bf r} = (r_x,r_y,r_z)$. In (\ref{NLS_per}) the spatial coordinates
are normalized to $\ell$, $\ell$ being a characteristic size of the
potential (say, the smallest of its periods), the time is mesured in
units of $2m\ell^2/\hbar$, and the amplitude of the order parameter is
normalized to the total number of atoms per unit volume
$\sqrt{{\mathcal N}/\ell^3}$. Then the nonlinearity coefficient $\chi$ is
given by $\chi=8\pi {\mathcal N}a_s/\ell$, where $a_s$ is the s-wave
scattering length. The potential $V({\bf r})$ is assumed, for the sake of
simplicity, to be separable, i.e. of the form
$V({\bf r})=\sum_{j} V_j(r_j), \ j=x,y,z$ (which corresponds to the majority
of experimental settings), and periodic in each of the spatial directions:
$V_j(r_j)=V_j(r_j+a_j)$, with $a_j$ the period in the direction
$r_j$ (in accordence with the accepted scaling $a_j\gtrsim 1$).
For convenience, the equation (\ref{NLS_per}) is considered subject
to periodic boundary conditions $\psi({\bf r})=\psi(r_x+L_x,r_y,r_z)$,
etc., where $L_j=N_j a_j$ with $N_j$ and $L_j$ respectively,
the number of primitive cells and the length of the system
in the direction $r_j$. The theory is developed for the small amplitude
limit, when the multiscale analysis is applicable. Hence, we look for
a solution to equation (\ref{NLS_per}) in the form
\begin{equation} \label{expansion}
  \psi=\epsilon\psi_1+\epsilon^2\psi_2+\epsilon^3\psi_3+\cdots
\end{equation}
where the $\psi_j$ are functions of the scaled independent variables
$\tau_p=\epsilon^p t$, ${\bf \xi}_{p}=\epsilon^p {\bf r}$, $p=0,1,2,...$,
with $\epsilon$ a small parameter to be specified later. Denoting with
$\omega_{\alpha_j}(q_j)$, and
$\Phi_{\alpha_j}(r_j)\equiv|\alpha_j,q_j\rangle$, the eigenvalues
and eigenfunctions of the periodic operators
$\cL_{r_j}=-\partial^2_{r_j}+V_j(r_j)$, we have
that the solution to a linear part of the equation (\ref{NLS_per}),
$\cL \psi=0$, with $\cL= i \partial_t-\sum_{j} \cL_{r_j}$,
can be written in the form
$|m_{x} m_{y} m_{z}\rangle =
\prod_{j} \Phi_{m_j}(r_j) e^{i\omega_{\alpha_j}(q_j) t}$, with
$\Phi_{m_j}(r_j)$ Bloch states of the corresponding 1D linear
problems. Here $m_j$ denotes the couple of quantum numbers
$\{\alpha_j,q_j\}$, with $\alpha_j$ the band index and $q_j$ the
component of the wave vector in the $j$ direction (note that the
imposed boundary conditions obviously imply that
$q_{j} \equiv q_{j,n}= \frac{2\pi}{L_j} n$ so that the extension of the
Brillouin zone (BZ) in the $j$ direction is $[-\pi/a_j,\pi/a_j]$).

Substituting the equation (\ref{expansion}) into (\ref{NLS_per}), and
collecting terms of equal powers in $\epsilon$, one arrives at the set
of equations $ \cL\psi_n={\cal M}_n $, with
$$
{\cal M}_1=0, \qquad
{\cal M}_2=-i\partial_{\tau_1}\psi_1- 2\nabla_0\nabla_1\psi_1,
$$
$$
{\cal M}_3 =-i\partial_{\tau_2}\psi_1 -
             i\partial_{\tau_1}\psi_2-\Delta_1\psi_1 -
             2\nabla_0 \cdot \nabla_1 \psi_2
            -2\nabla_0 \cdot \nabla_2 \psi_1 + \chi|\psi_1|^2\psi_1,
$$
where $\nabla_p$ denotes the gradient with respect to ${\bf \xi}_p$.

Since we are interested in instabilities of the condensate
wavefunction, we investigate the influence of the nonlinear term
in the equation (\ref{NLS_per}) on the Bloch states of the underlying
linear problem. To this end we take as starting point in the
expansion (\ref{expansion}), a modulated state of the form
\begin{equation} \label{order_1}
  \psi_1={\cal A}({\bf \xi}_1,...; \tau_1,...)
  e^{-i\omega_0\tau_0}|m_{0x} m_{0y} m_{0z} \rangle,
\end{equation}
with $\omega_0\equiv \sum_j\omega_{\alpha_{0,j}}(q_j)$
(the subscript zero refers to the chosen band, below we consider the two
lowest ones). Then the first order equation is automatically satisfied by
$\psi_1$, while the equation of the second order can be solved in the form
\begin{equation} \label{order_2}
  \psi_2=\sum_{\bal}\!^{^\prime} {\cal B}_{\bal}
  e^{-i\omega_0\tau_0} |\alpha_x,q_{0,x};\alpha_y,q_{0,y};
  \alpha_z,q_{0,z}\rangle,
\end{equation}
where the prime denotes ${\bf \alpha} \neq {\bf \alpha}_0$ in the
sum and have taken into account that the terms with  ${\bf q}\neq{\bf q}_0$
give zero contribution. Analysis similar to that of reference \cite{ks}
shows that $\cA=\cA({\bf R};\xi_2...;\tau_2,...)$ with
${\bf R}={\bf \xi}_1-{\bf v}\tau_1$ and ${\bf v}=-\langle \alpha_{0x}
\alpha_{0y} \alpha_{0z}|2i\nabla |\alpha_{0x} \alpha_{0y}
\alpha_{0z} \rangle$ is the group velocity of the carrier wave.
The coefficients ${\cal B}_{\bal}$ are found as
\begin{eqnarray} \label{B}
  {\cal B}_{{\bf \alpha}} = \frac{
  \Gamma^{(y,z)}_{\alpha_x,\alpha_{0,x}} \partial_{x_1} \cA +
  \Gamma^{(x,z)}_{\alpha_y,\alpha_{0,y}} \partial_{y_1} \cA +
  \Gamma^{(x,y)}_{\alpha_z,\alpha_{0,z}} \partial_{z_1} \cA }
  {\omega_0-\omega_{\bal}(\bq_0)},
\end{eqnarray}
where
$\omega_{\bal}(\bq_0)=\sum_j\omega_{\alpha_{j}}(q_{0j})$,
$\Gamma^{(y,z)}_{\alpha_x,\alpha_{0,x}}=
-\langle \alpha_x,q_{0,x}|2\partial_{x_0}|
\alpha_{0,x},q_{0,x}\rangle\delta_{\alpha_y,\alpha_{0,y}}
\delta_{\alpha_z,\alpha_{0,z}}$ (other coefficients $\Gamma$ are
obtained by cyclic permutations of $x,y,z$). Finally, considering
the orthogonality of ${\cal M}_3$ to $|m_{0x} m_{0y}
m_{0z} \rangle$ we obtain the following 3D NLS for the slowly varying
envelope
\begin{eqnarray} \label{NLS}
  i\frac{\partial \cA }{\partial \tau_2} + \frac{1}{2} \sum_{j=x,y,z}
  {{\bf M}}_{\alpha_j,jj}^{-1} \frac{\partial^2 \cA}{\partial R_{j}^2} -
  \tilde{\chi}|\cA|^2\cA=0,
\end{eqnarray}
where we assumed $\cA$ not depending on ${\bf \xi}_2$, and
introduced the inverse of the effective mass tensor
\begin{equation} \label{effmass}
\frac{1}{2}{{\bf M}}_{\alpha_j,jj}^{-1} = 1+
\sum_{\alpha_x}\frac{|\Gamma_{\alpha_x\alpha_{0,x}}^{(y,z)}|^2}
{\omega_{\alpha_x}(q_{0,x})-\omega_{\alpha_{0,x}}(q_{0,x})} =
\frac{1}{2}\partial_{ q_j}^2\omega_{\alpha_j} ({\bf q}),
\end{equation}
and the effective nonlinearity
\begin{equation} \label{effnonl}
\tilde{\chi}=\chi\prod_{j=x,y,z}\int_{0}^{L_j}
|\Phi_{m_{0j}}|^4\,dr_j.
\end{equation}

Now we are at the point to discuss the small parameter. To
simplify, we consider a cubic box with $a_j=\ell$ and $L_j=L$.  On
the one hand, as it was mentined above, the physical order parameter
is normalized to the total number of atoms ${\cal N}$, while the formal
wave function $\psi$ must be normalized to one: $\int|\psi|^2d{\bf r}=1$.
On the other hand all parameters in the equation (\ref{NLS}) must be of
order one. Next we notice the oscillatory character of the Bloch
functions, in a general situation leads to the fact that the integrals in
the last expression (\ref{effnonl}) for $\tilde{\chi}$  has a numerical
smallness (see e.g. the examples below). Then, taking into account that
$\chi=8\pi {\cal N}a_s/\ell$ one can define $\epsilon={\cal N}a_s\ell^2/L^3$.
Consider now a condensate with ${\cal N}=10^5$ of $^{87}$Rb atoms
($a_s\approx 5.5$ nm) homogeneously
distributed over a cubic box with $L=100\,\mu$m having $N_i=100$ cells in
each direction (respectively $\ell=1\,\mu$m). Then we compute
$\epsilon \approx 0.014$. A physical situation when $\epsilon$ is
not too small (say in experiments \cite{grainer2} it can be
identified as $\epsilon \approx 0.257$) the multiple scale
expansion, strictly speaking, is not valid. For this reason below
we employ the numerical simulations, which, however, clearly
illustrate that the small amplitude limit gives remarkably good
estimates for the characteristic scales of the problem and allows
one to understand the symmetry of the developed patterns.

Let us analyze the stability problem within the framework of
equation (\ref{NLS}), i.e. look for a solution of the form
(\ref{order_1}) with $ \cA=\left(\rho+ae^{i(\Omega \tau_2-{\bf
K}{\bf R})}+ be^{-i(\Omega t_2-{\bf K}{\bf
R})}\right)e^{-i\rho^2\tau_2} $, where $|a|\,,|b|\ll \rho$. This
solution is modulationally unstable if
\begin{equation} \label{instab}
  Z(Z+4\tilde{\chi}\rho^2)<0,\qquad
  Z=\sum_{j=x,y,z}{{\bf M}}_{\alpha_j,jj}^{-1} K_j^2 ,
\end{equation}
where the modulational instability is understood in the sence of a plain
wave instability with respect to small modulations of its amplitude.
Below in section \ref{numerics} we consider the modulational instability
of solutions to equation (\ref{NLS_per}) in the form of Bloch
states induced by periodic small amplitude and long-wavelength
perturbation.

\section{Energy band structures}
\label{bands}

\begin{figure}[htb]
\centerline{
\includegraphics[width=0.7\textwidth,clip]{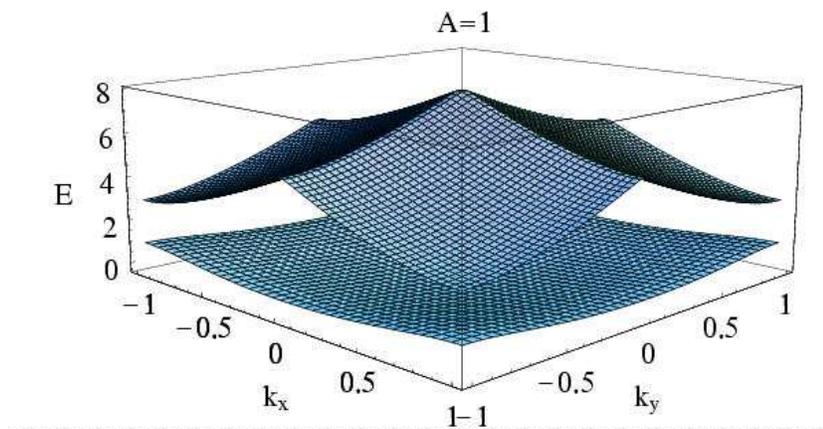}
}
\caption{The first two bands of the 2D
separable Mathieu potential of amplitude $A=1$ and lattice
constants $a=b=\pi$ plotted in the reduced zone scheme.}
\label{f1}
\end{figure}

In the previous section we assumed the knowledge of
the Bloch states and energy bands of the underlying linear
Schr\"odinger problem. For generic multidimensional potentials
this could be a quite difficult problem to solve. To avoid
difficulties we shall restrict here to the case of separable
potentials of trigonometric form i.e. we take $V({\bf r})=\sum_{j}
V_j(r_j), \ j=x,y,z$ with $V_j(r_j)$ given by
\begin{equation}
 V_j(r_j)=A \cos(\kappa_j r_j).
\label{mathiew}
\end{equation}
Here $A$ is a constant fixing the depth of the lattice and $2
\pi/\kappa_j$ the periodicity in the $r_j$ directions. In the
following we fix $\kappa_j=2$ for all $j$ so that the potential
will be a superposition of identical one dimensional Mathieu
potentials, and the corresponding Bravais lattice will have simple
cubic symmetry. The band structure and the Bloch states of the
linear Schr\"odinger equation are then obtained as
\begin{equation}
  E_{\alpha}(\bf{k}) = \sum_{j} \epsilon_{\alpha}(k_j),\;\;
  \Psi_{\alpha}({\bf{k}},{\bf{r}})=\prod_{j} \varphi_{\alpha}
  (k_j,r_j),
\end{equation}
where $\alpha$ denotes the band index and $\epsilon_{\alpha}(k_j)$
and $\psi_{\alpha} (k_j,r_j)$ are the eigenvalues and
eigenfunctions of the Mathieu equation
\begin{equation}
  -\frac{d^{2}\varphi _{k,\alpha }}{dx^{2}}+A \cos(2 x) \varphi
  _{k,\alpha }=E_{\alpha }(k)\varphi _{k,\alpha }. \label{mathieu}
\end{equation}
This equation can be transformed, making use of the expansion of
the wave function into momentum eigenfunctions (Fourier
expansion), into a tridiagonal problem whose solutions can be
obtained with high accuracy by means of continued fractions. As an
example of these calculations we report in figure \ref{f1} the
first two energy bands of the 2D separable Mathieu potential, in the
case $A=1$ (note that with the choice $\kappa=2$ the potential
has periodicity $\pi$ in both directions, so that the BZ is a square of
size 2).

\begin{figure}[htb]
\centerline{
\scalebox{0.25}{\includegraphics{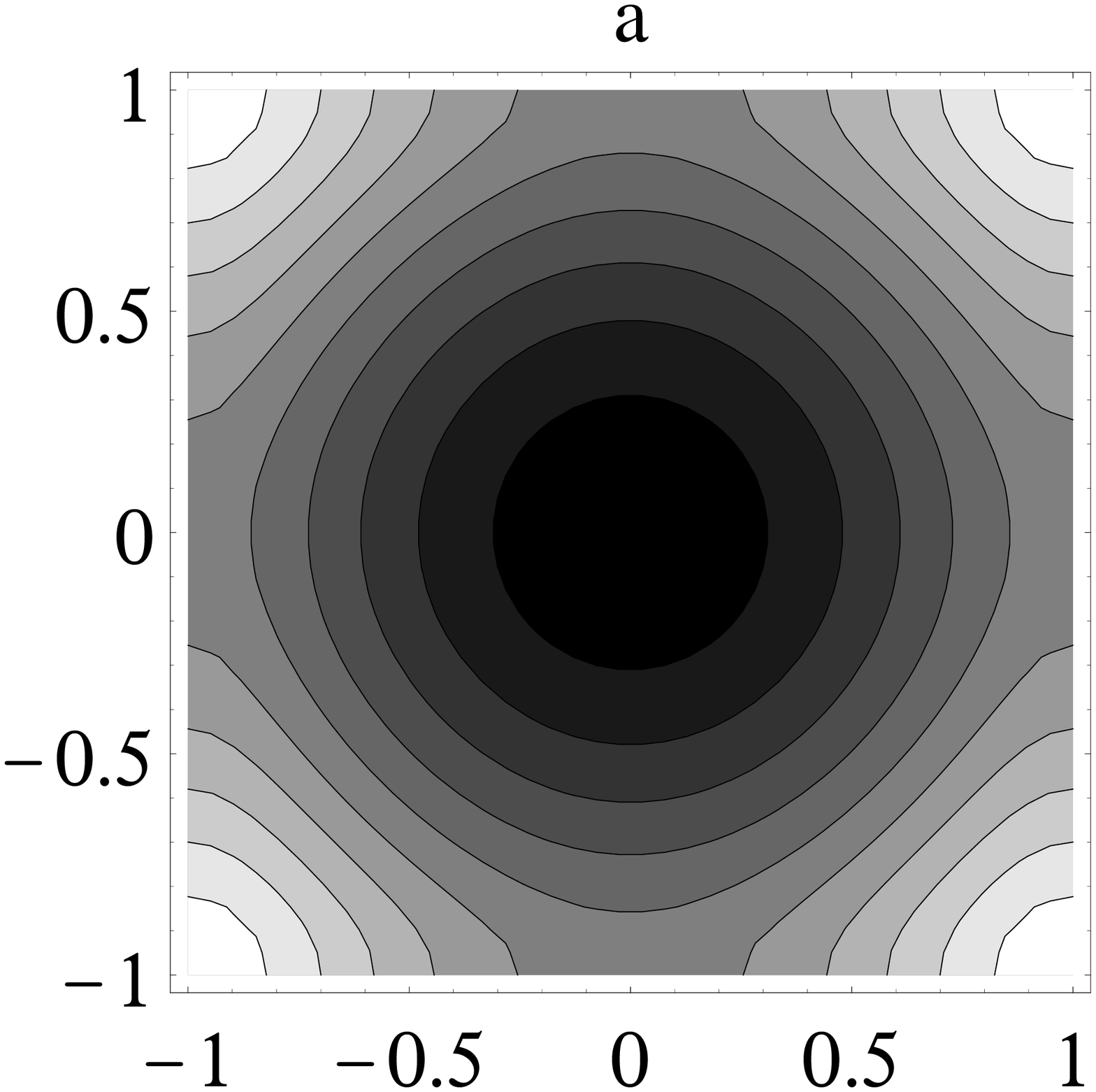}} \quad
\scalebox{0.25}{\includegraphics{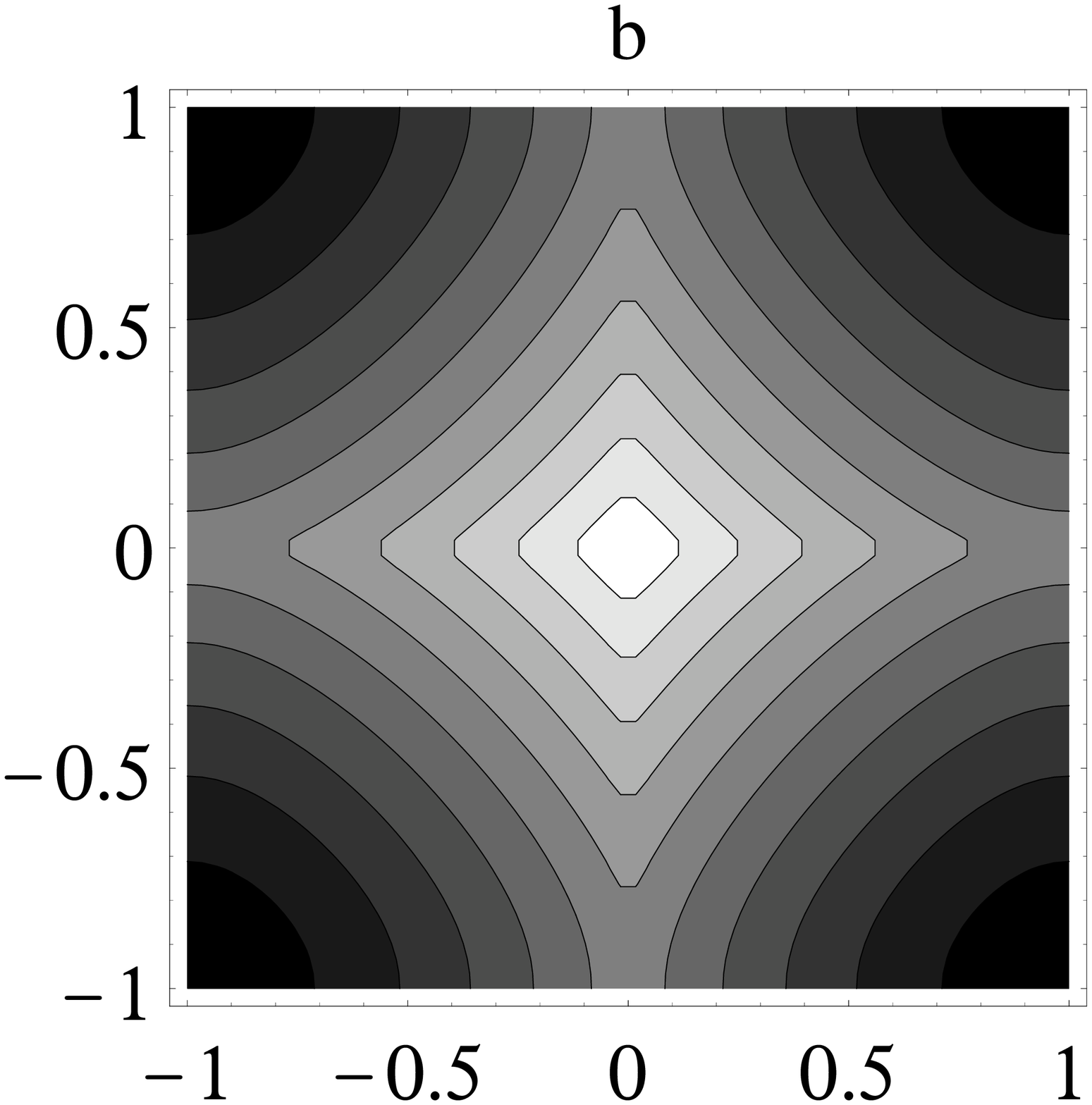}}
}
\caption{Sections of constant energy for the lower (a) and
upper (b)  bands in figure~\ref{f1}. The white regions denote
areas of higher energy.}
\label{f2}
\end{figure}

\begin{figure}[htb]
\centerline{
\scalebox{0.25}{\includegraphics{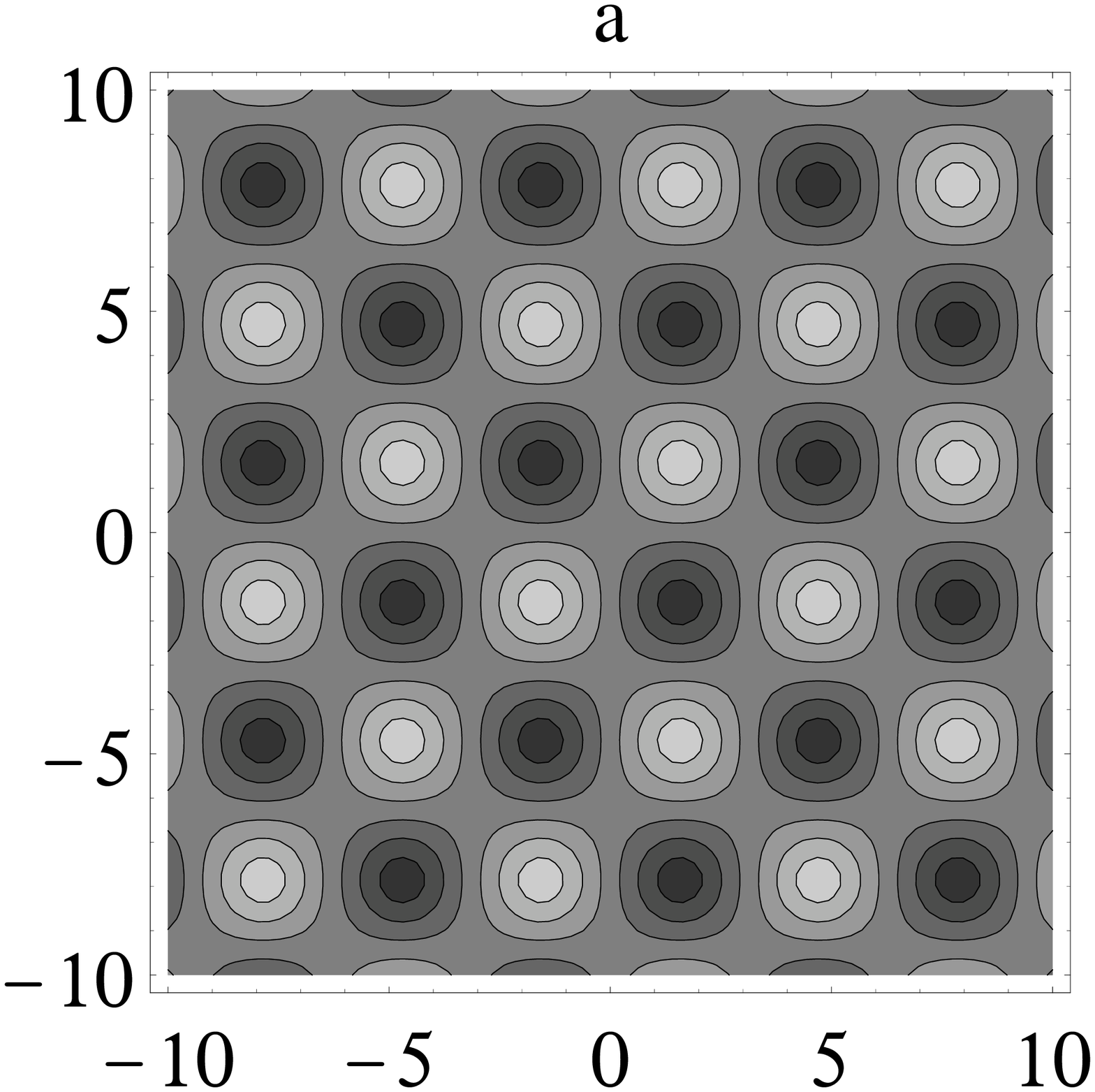}} \quad
\scalebox{0.25}{\includegraphics{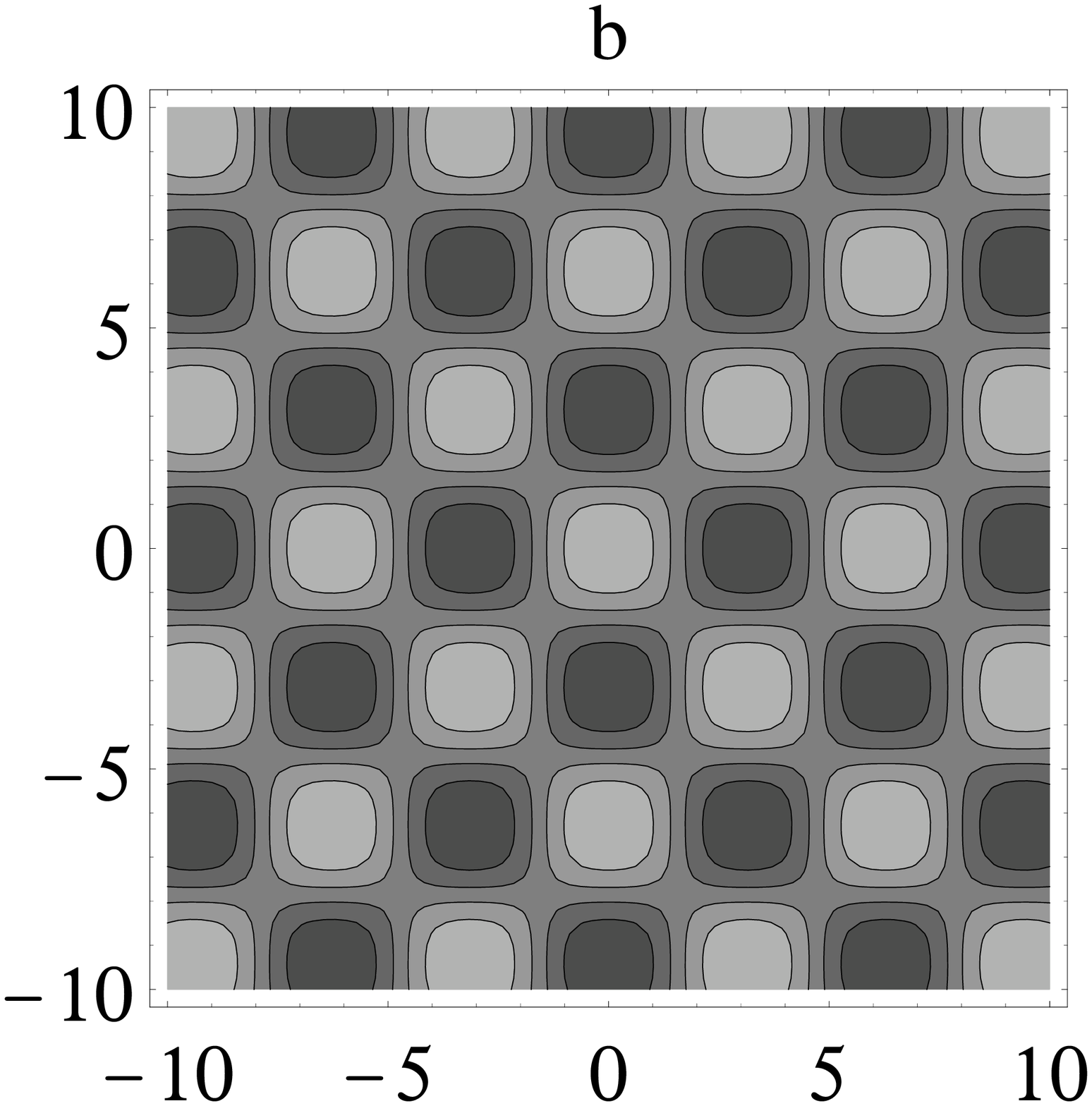}}
}
\caption{Contour plots of the Bloch states at
the corner $k_x=1, k_y=1$ of the BZ for the lower (a)
and upper (b) band of figure 1. The wavefunction in (a) leads to
soliton formation trough modulational instability while the one in
(b) is modulationally stable.}
\label{f3}
\end{figure}

\begin{figure}[htb]
\centerline{
\scalebox{0.25}{\includegraphics{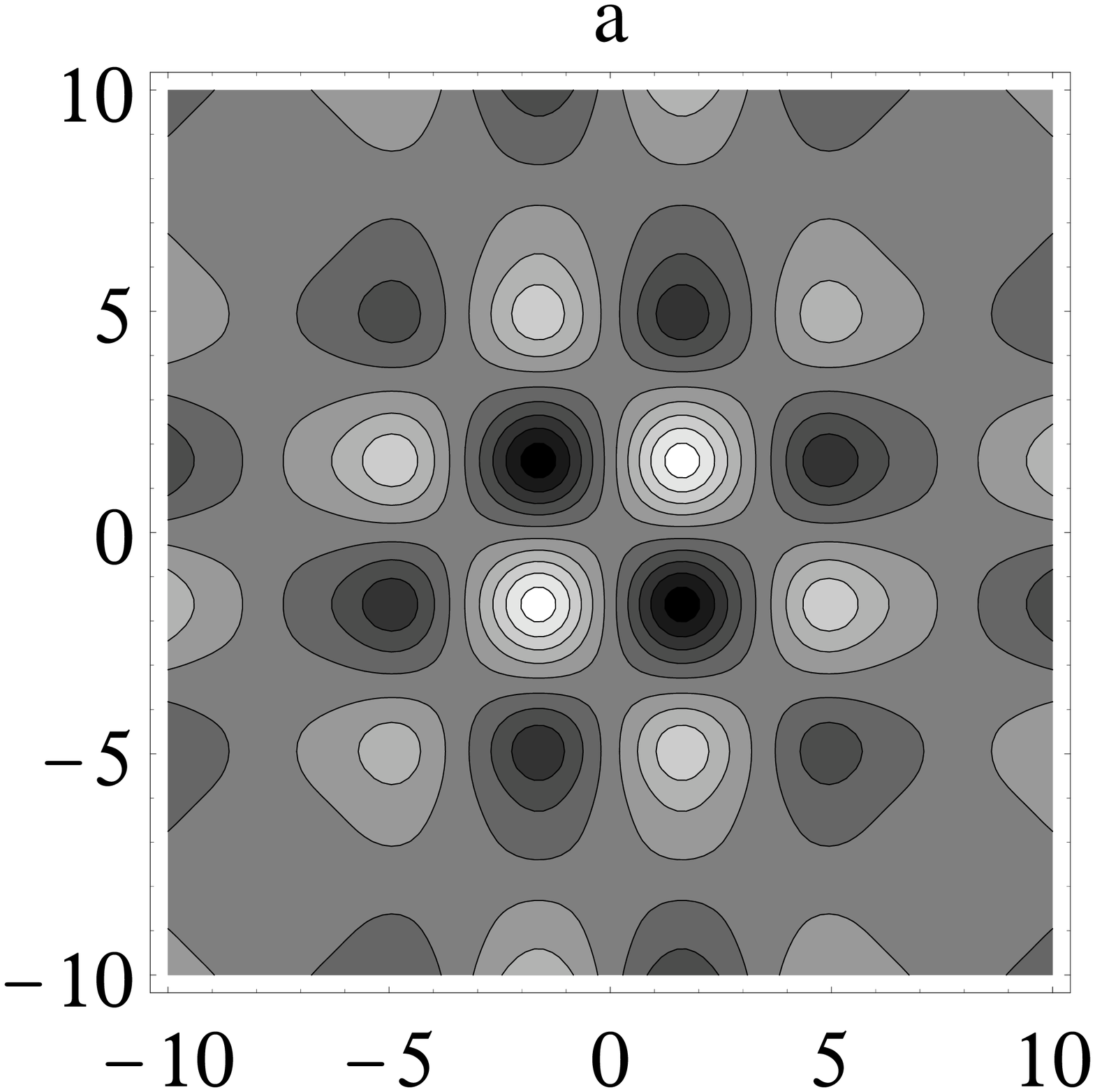}} \quad
\scalebox{0.25}{\includegraphics{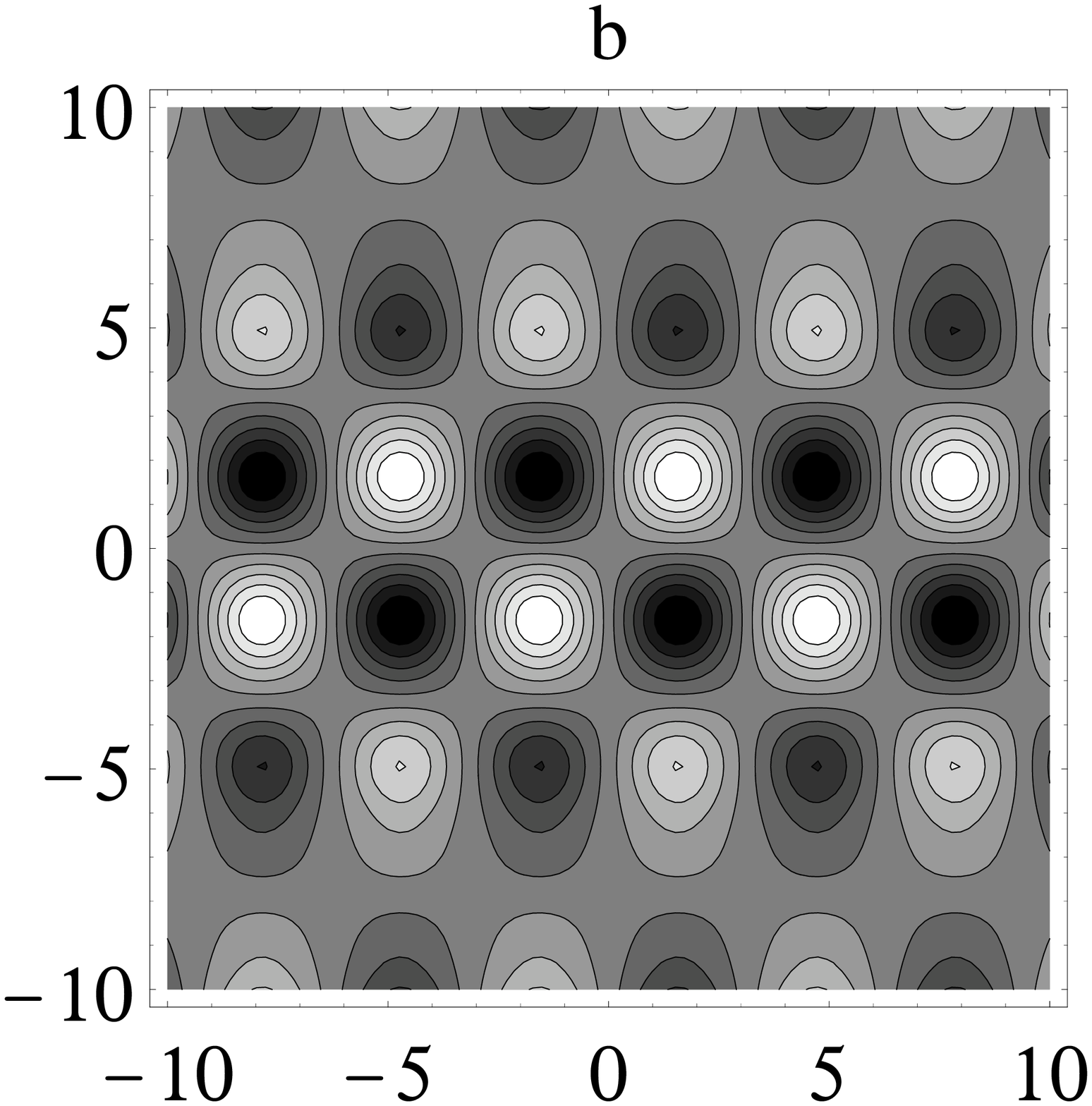}}
}
\caption{Contour plots of Bloch states of the lower band along
lines of symmetry of the BZ. Case (a) refers to
$k_x=0.8, k_y=0.8$ while (b) to $k_x=1.0, k_y=0.8$.}
\label{f4}
\end{figure}

\begin{figure}[ht]
\centerline{
\scalebox{0.28}{\includegraphics{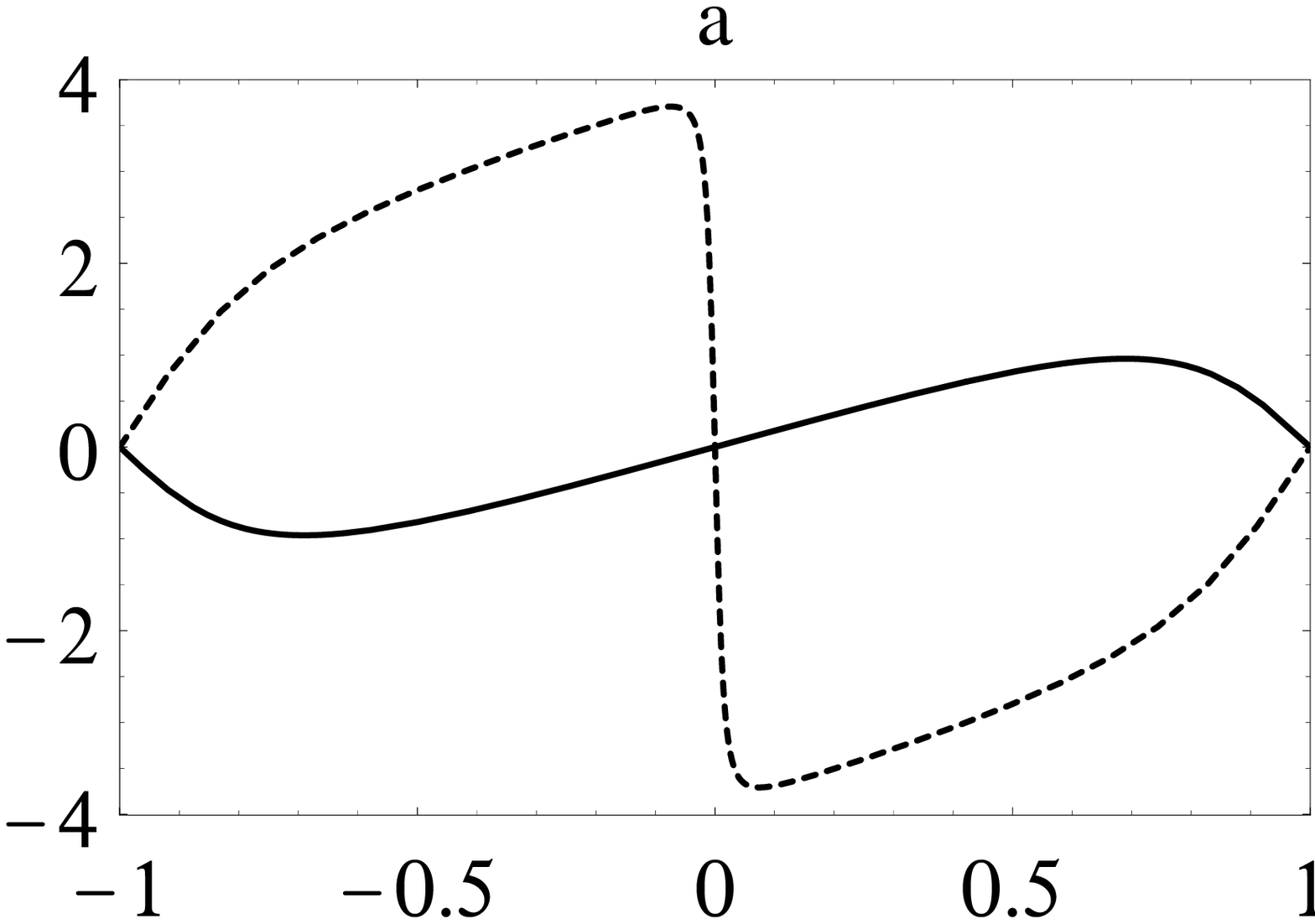}} \quad
\scalebox{0.28}{\includegraphics{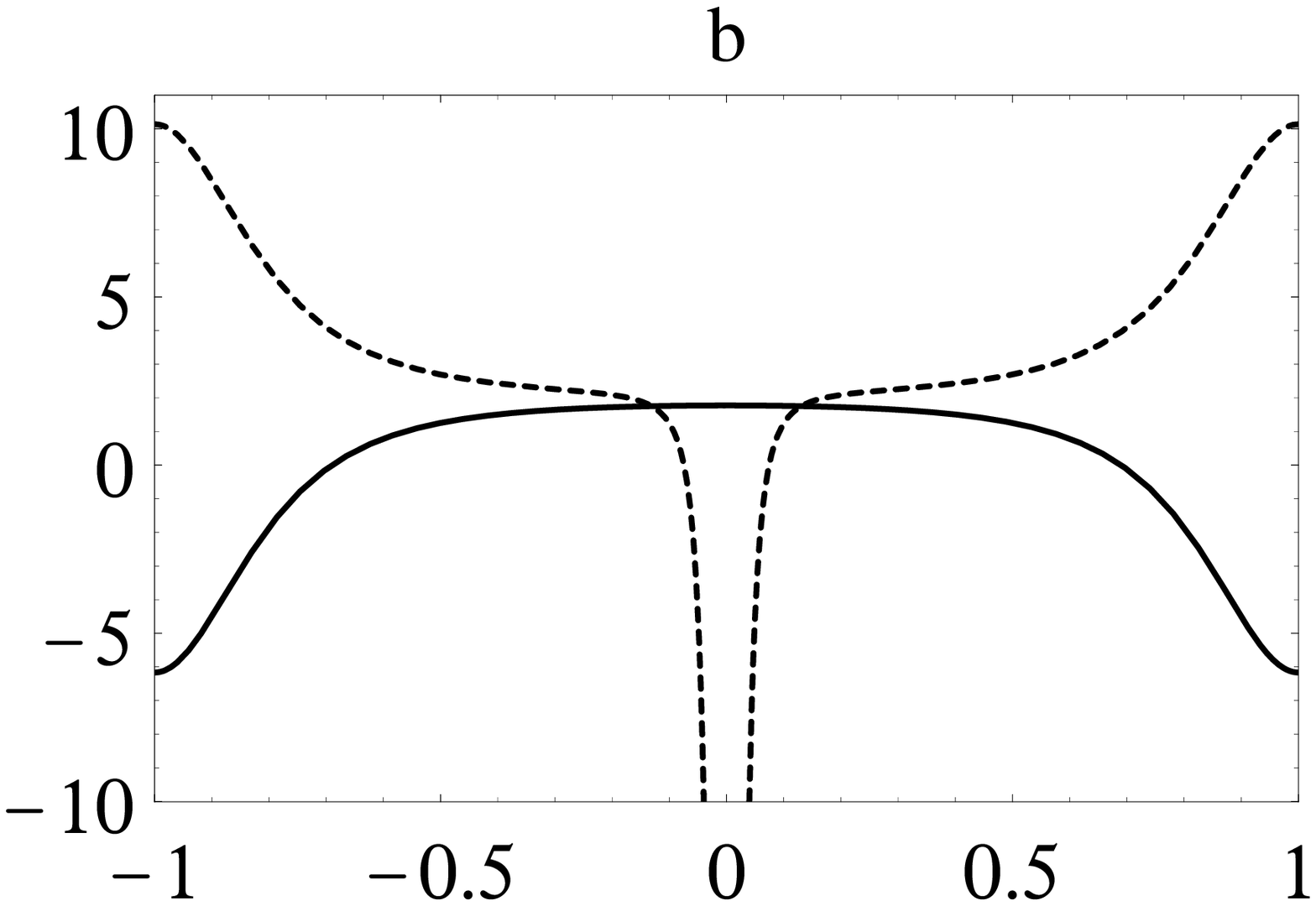}}
}
\caption{The group velocity (a) and the
reciprocal of the effective mass $\omega_{xx}$ (respectively
$\omega_{yy}$) (b), as a function of $k_x$ (respectively $k_y$), for
arbitrary $k_y$ (respectively $k_x$) in the BZ. The continuous and broken
lines refer, correspondingly, to the first and second band of figure
\ref{f1} (notice from (a) that there is no singularity for
$\omega_{xx}$ (or $\omega_{yy}$) in the origin.}
\label{f5}
\end{figure}

In figure \ref{f2} we also show the sections of constant energy
for the bands depicted in figure \ref{f1}. By changing the
amplitude of the potential the band structure will change, and the
bands become more flat and more separated as the amplitude of the
potential is increased.
In figures \ref{f3},\ref{f4} we show the
contour plots of Bloch states at different points in the BZ.
We remark that, due to the separability of the potential, both the
derivative and the curvature of the band are independent on $k_y$
(respectively $k_x$) for fixed values of $k_x$ (respectively $k_y$) in
the BZ. This is seen in figure \ref{f5}, where the group velocity
and the components of the reciprocal mass tensor are reported  for
the bands in figure \ref{f1}.
Similar calculations can be easily done for the 3D separable
Mathieu potentials. The analysis can be extended to lattices with
rectangular or tetragonal symmetry, as well as, to more general
separable potentials such as the ones considered in \cite{Belokolos}.

In the following sections we shall use these results to compare
numerical studies of the instability of Bloch states in presence
of nonlinear interactions with the prediction (equation (\ref{instab}))
of the previous section.

\section{Numerical simulations}
\label{numerics}

The linear stability analysis, described in section \ref{model} gives
the growth rates and spectra for the modulational instability. This
information appears to be sufficient to predict the spatial arrangement and
symmetry of the emerging soliton-like structures. However, to gain more
insight into the development of modulational instability one has to
recourse to numerical simulations.

For the numerical study we have used the potential
$V_j(r_j)=A \cos (\kappa r_j)$ for $j=x,y,z$, which is motivated by the
recent experiments \cite{grainer1,grainer2}. Then in the above formulas
the terms corresponding to $j=x$, $j=x,y$, and $j=x,y,z$ are retained,
respectively for 1D, 2D and 3D optical lattices. Also in the case at hand
${{\bf M}}_{\alpha,xx}^{-1}$, ${{\bf M}}_{\alpha,yy}^{-1}$, and
${{\bf M}}_{\alpha,zz}^{-1}$  have the same functional dependence on the
arguments, which means that they coincide when $q_x=q_y=q_z$.
Numerical solution of the equation (\ref{NLS_per}) has been performed
by the operator splitting procedure using multi-dimensional
fast Fourier transform \cite{numrecipes}. The spatial domain
$x,y,z \in [-{L \over 2}..{L \over 2}]$ (i.e. $L_x=L_y=L_z=L$)
was represented by an array of 128 x 128 x 128 points. For 1D and 2D cases
the results were checked by increasing the number of grids (512 and
256 x 256 respectively), which showed no qualitative difference.
The time step was $\delta t = 0.001$.
To be specific, we concentrate on the case of
positive scattering length, $\chi=1.0$, choose $\kappa=2.0$
(i.e. $a_x=a_y=a_z=\pi$), and $\rho=0.5$.

\subsection{1D optical lattice}
\label{1dcase}

Basic features of the development of modulational instability and formation
of soliton-like excitations in effectively 1D optical lattice was described
in \cite{ks,alfimov}. Below we extend the parameter values, which can
lead to formation of qualitatively different types of localized excitations.

\begin{figure}[htb]
\centerline{\hspace*{1.5cm}
\includegraphics[width=6.0cm,height=4.5cm,clip]{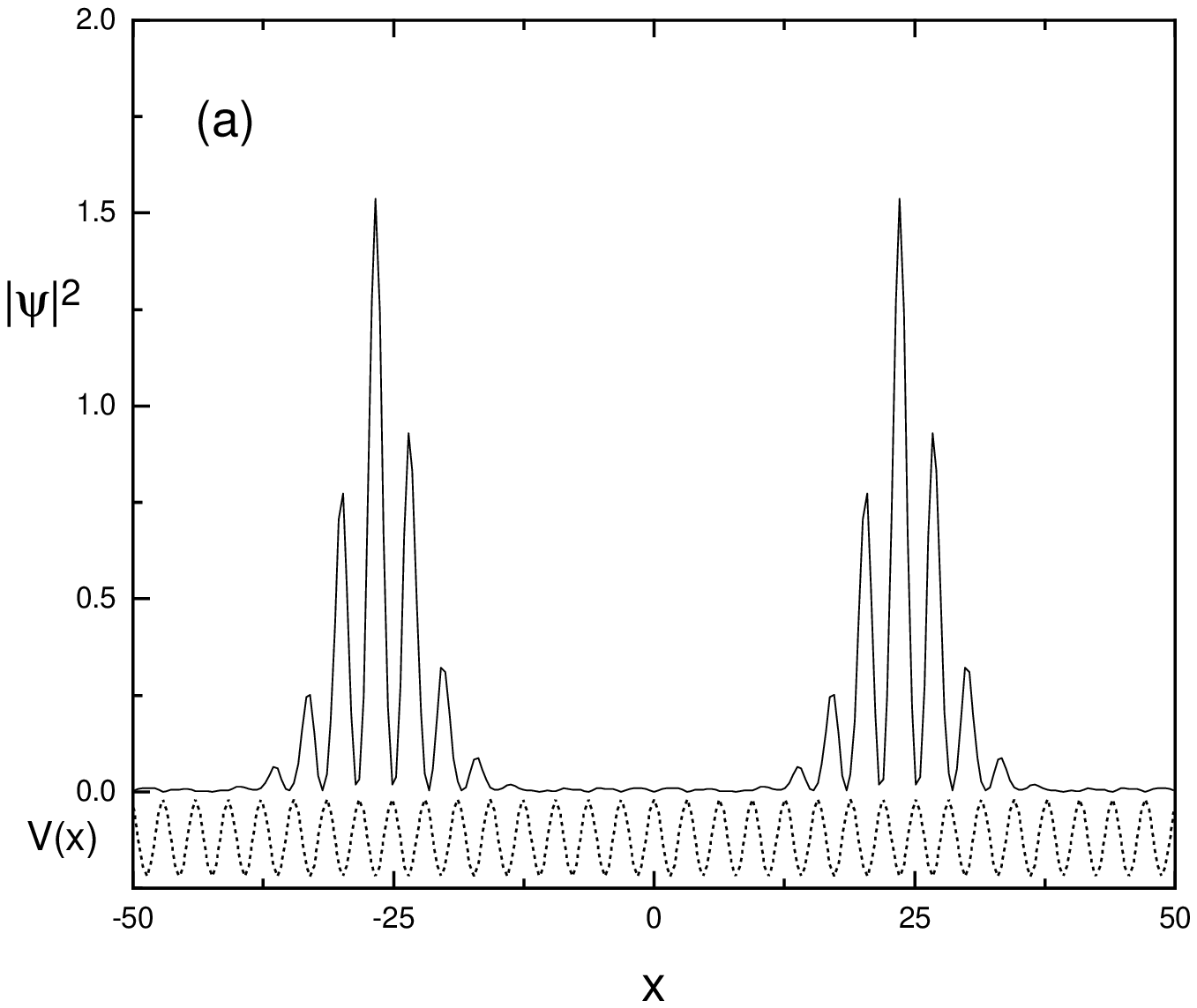} \qquad
\includegraphics[width=6.0cm,height=4.5cm,clip]{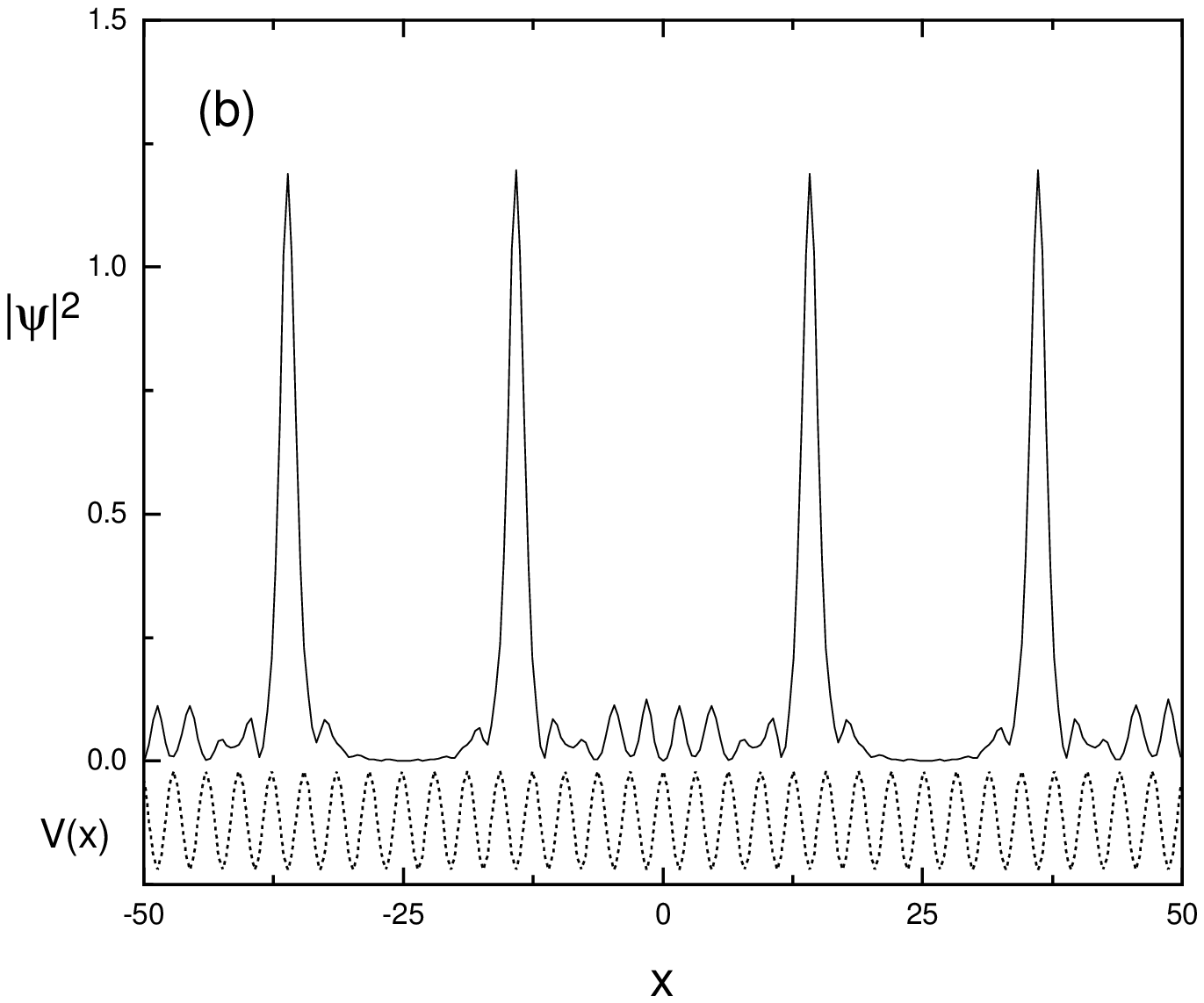}
} \caption{Two types of soliton-like excitations in arrays of BECs
confined to 1D optical lattice, emerging due to the modulational
instability of the initial waveform $\psi(x,0)=0.5 \sin(x)$ at
$t=0$. The lower curve represents the optical lattice potential
$V(x)=A \cos(\kappa x)$. (a) Envelope solitons, occupying few
lattice sites, are formed at $t=400 \ $ when the nonlinear atomic
interactions are weak $\chi = 0.2$. (b) Intrinsic localized modes,
which fit a single lattice site, are formed at $t=70$ for strong
nonlinearity $\chi = 1.0$. Parameter values $A=1.0, \ \kappa=2.0,
\ L=32 \pi$. The initial waveform is perturbed by
$\delta \psi (x) = 0.01 \sin(0.125 x)$ } \label{f6}
\end{figure}

The coefficient of nonlinearity $\chi$ in equation (\ref{NLS_per})
is an important parameter which determines such a property of BEC
as the macroscopic quantum self-trapping \cite{smerzi,milburn,raghavan}.
At strong nonlinearity the tunneling of atoms between adjacent wells of the
optical lattice is suppressed, dispite the significant population
imbalance, due to macroscopic quantum self-trapping effect. This
property affects the development and further evolution of
spatially localized excitations in BEC arrays. In figure
\ref{f6} we report two types of soliton-like excitations,
developed in 1D BEC arrays at weak and strong nonlinearities. All
remaining parameters, except $\chi$, are similar in these two
cases. Envelope soliton-like modes, which occupy few lattice sites
are formed at weak nonlinearity, while the intrinsic localized
modes, occupying a single lattice site are formed at strong
nonlinearity. The time required for development of these
excitations are also different.
The localized excitations represented in the figure \ref{f6} are
thin disks filled in with BEC atoms, where the atomic density is
much greater than in neighbouring array sites. The dynamics of
these excitations is governed by the 1D nonlinear Schr\"odinger
equation \cite{ks}, and for particular parameter settings they can
be stable \cite{alfimov}, or have very long (relative to duration
of experiments) recurrence times. The separability of the periodic
lattice potential in equation (\ref{NLS_per}) leads to similar
scenarios of the development of modulational instability also in
2D and 3D cases considered below.

\subsection{2D optical lattice}
\label{2dcase}

Now let us consider in more detail the development of soliton-like
excitations and the possibility to stabilize them in a 2D optical lattice.
A 2D optical lattice is formed by overlapping two laser standing waves
along the $x$ and $y$ axes, superimposed on a continuous BEC in a magnetic
trap. The condensate is then fragmented and confined in many narrow tubes
centered at lattice potential minima and directed along the $z$ axis.
As a result of modulational instability, the initial distribution of the
atomic density over the tubes in the optical lattice is changed.

In order to analyze the instability of initial waveforms we
consider Bloch states corresponding to different points of the BZ.
Let us consider the points ${\bf q}_0= (\pm 1,\pm 1)$ at the
boundary of the BZ (figure \ref{f1}). Then, restricting
consideration to the two lowest bands ($\alpha=1,2$), one can
distinguish three different cases:

\subsubsection{Case 1.}

\begin{figure}[htb]

\centerline{
\includegraphics[width=8.0cm,height=3.5cm,clip]{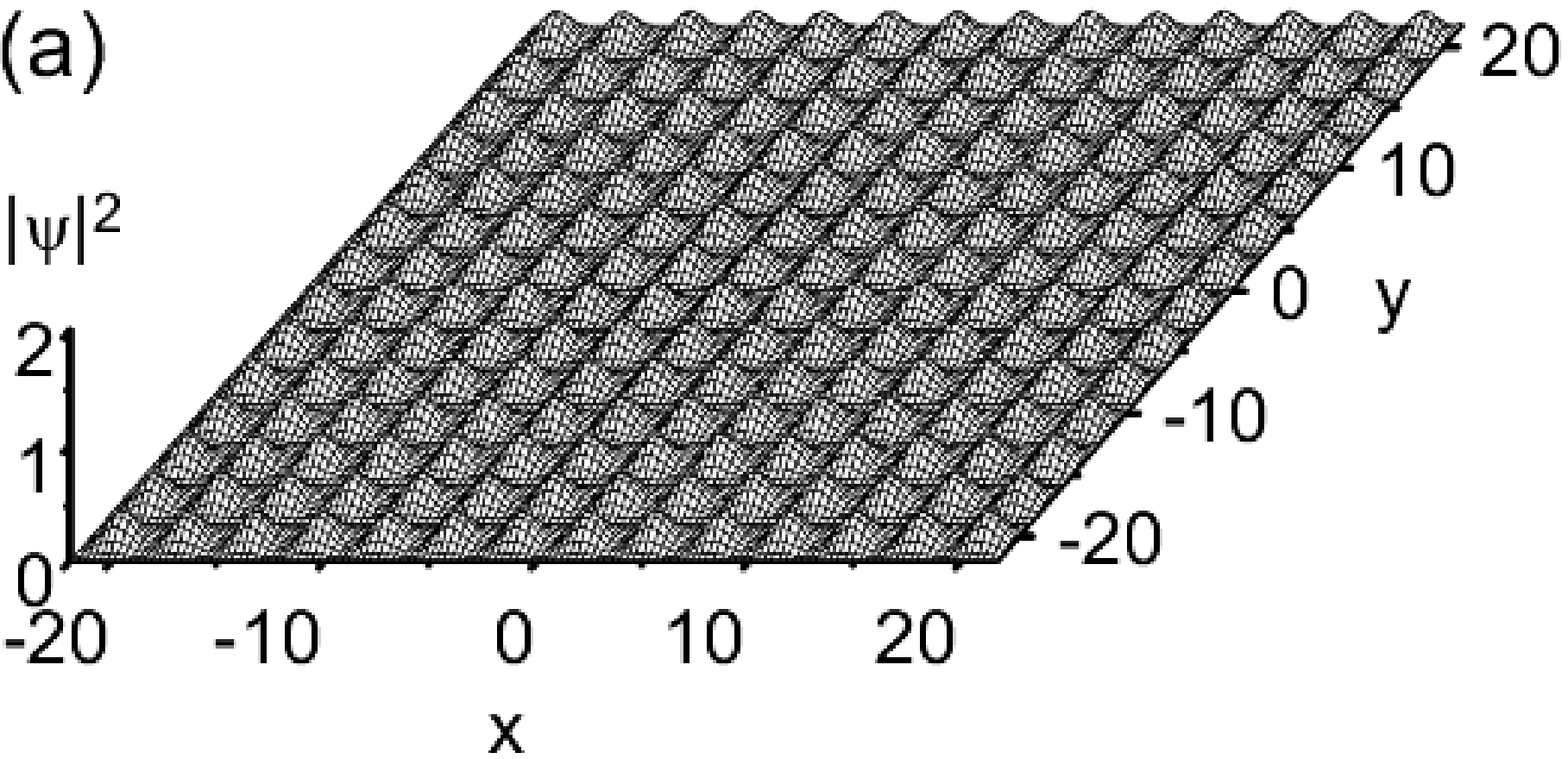}}
\vspace*{0.5cm}
\centerline{
\includegraphics[width=8.0cm,height=3.5cm,clip]{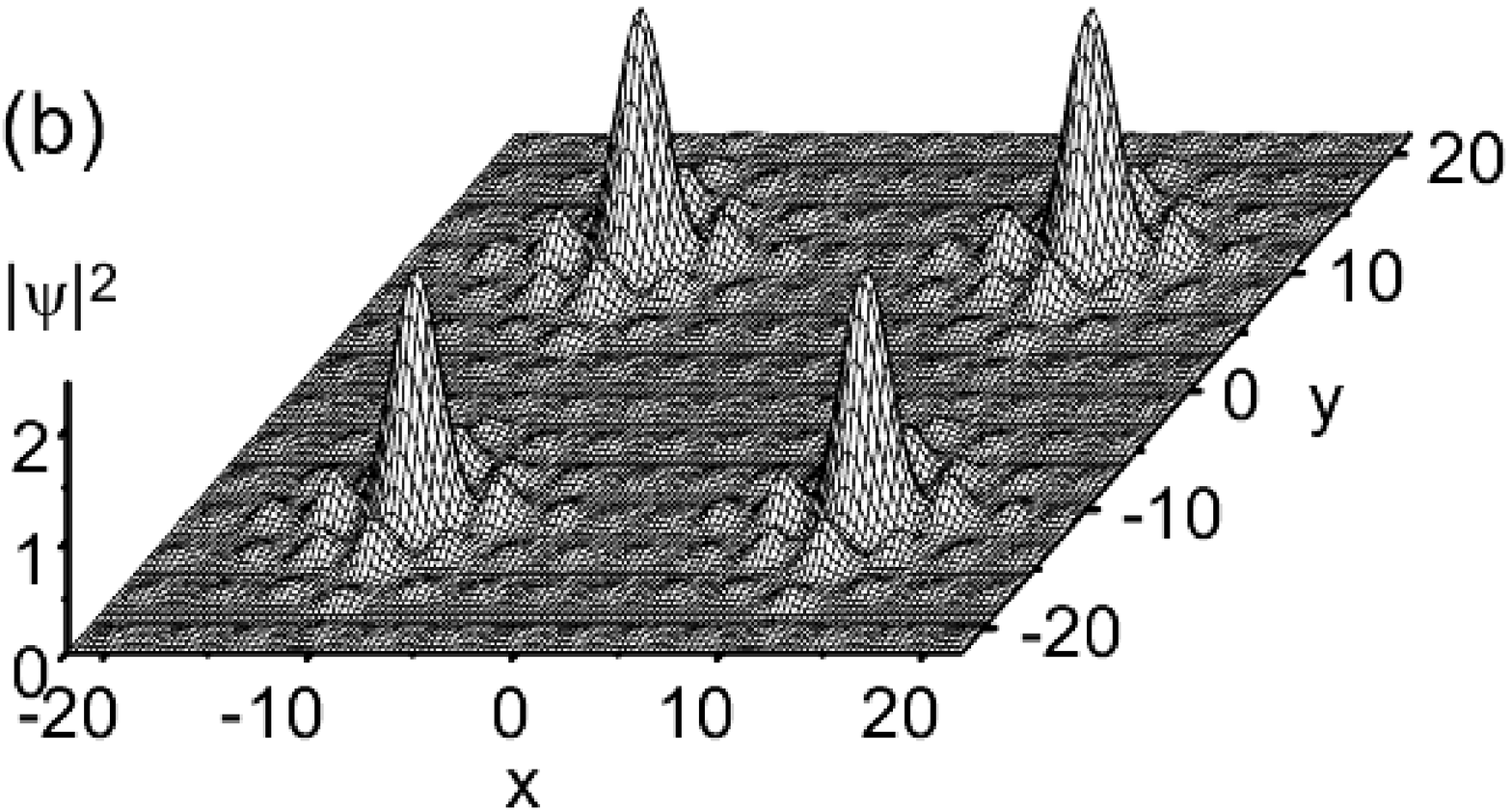}}
\vspace*{0.5cm}
\centerline{
\includegraphics[width=8.0cm,height=3.5cm,clip]{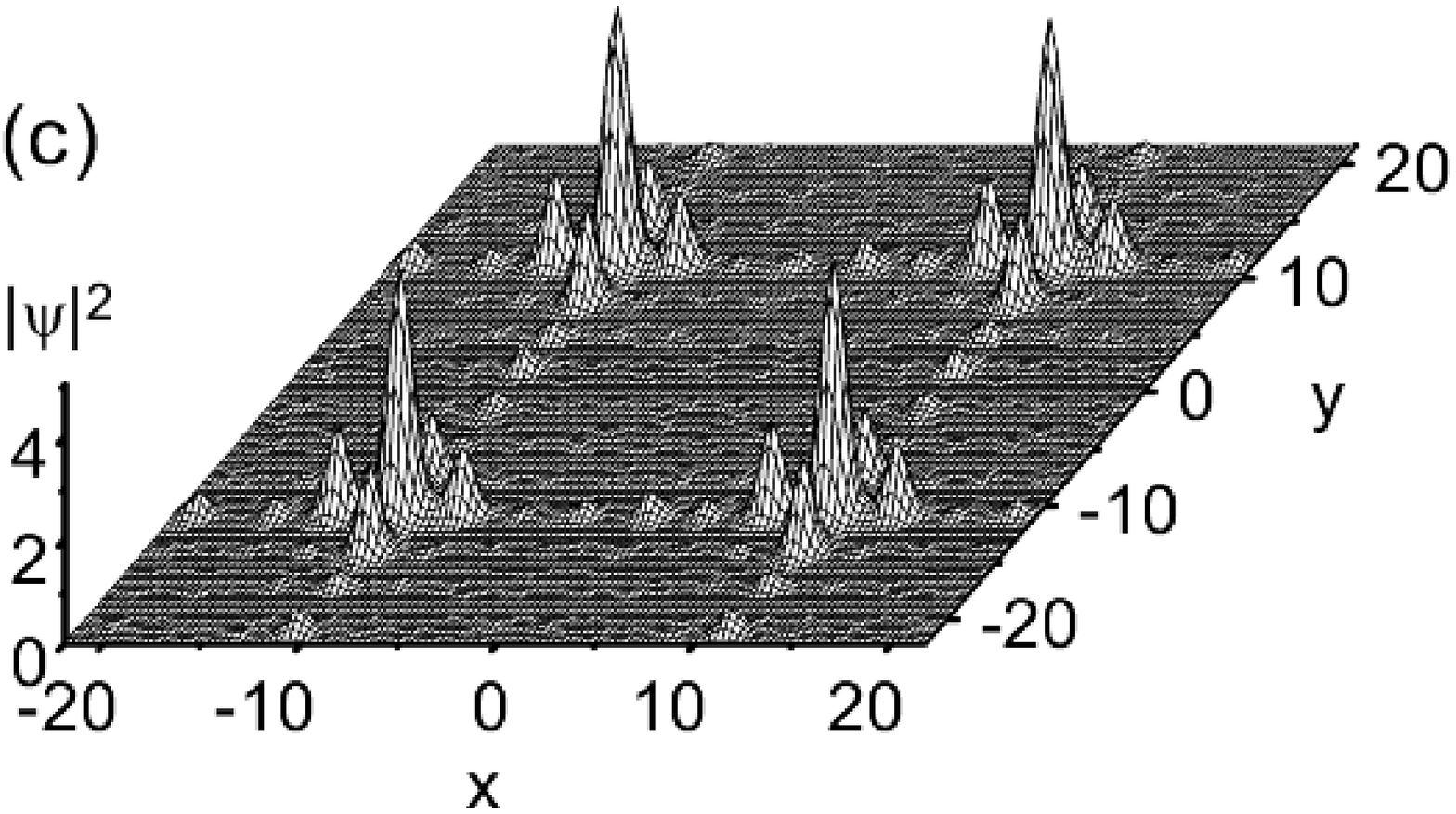}}
\vspace*{0.5cm}
\caption{The evolution of BEC atomic distribution in a 2D optical lattice
according to equation (\ref{NLS_per}) with
$A=1.0, \ \kappa=2.0, \ \chi=1.0, \ L=14\pi. \ $
(a) The initial waveform $\psi(x,y,0) = 0.5 \sin(x) \sin(y)$ at $t=0$.
(b) Formation of soliton-like excitations due to modulational instability
at t=85 when the initial waveform is perturbed by
$\delta \psi (x,y) = 0.01 \sin(0.125 x) \sin(0.125 y)$.
(c) The distribution (b) slightly narrows but remains stable for very
long times when the strength of the trap potential is adiabatically increased
up to $A=3.0$ during $85 < t < 90$. The snapshot (c) is shown
at $t=200$ (stability is verified up to t=1000). }
\label{f7}
\end{figure}

\begin{figure}[htb]
\vspace*{0.5cm}
\centerline{\includegraphics[width=8.0cm,height=3.5cm,clip]{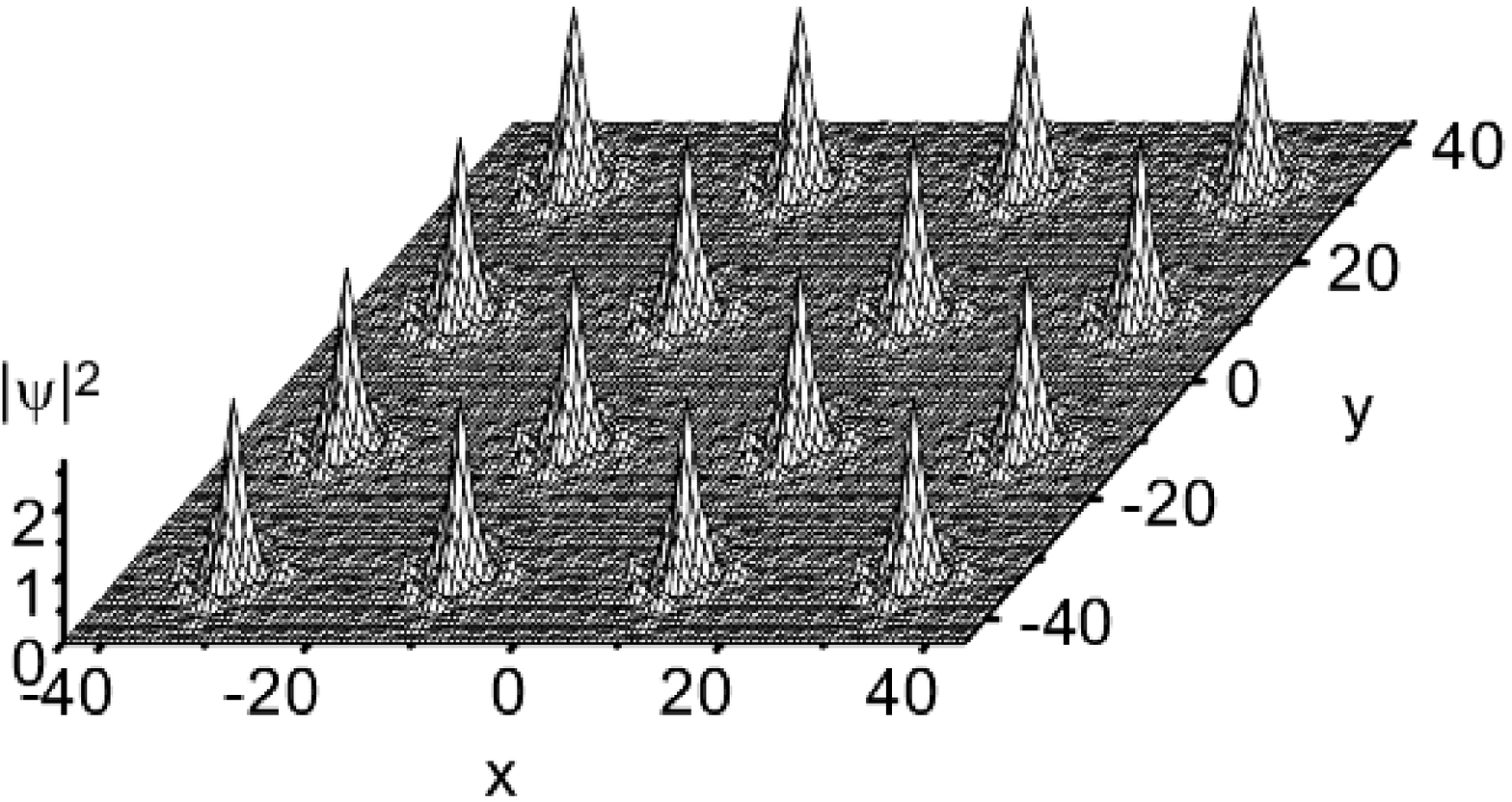}}
\vspace*{0.5cm}
\caption{Regular spatial structures with BEC in a 2D optical lattice,
emerged due to modulational instability. Parameter settings are similar
to those of figure \ref{f7}, except the domain size $L=28\pi$.}
\label{f8}
\end{figure}

Both eigenfunctions $|m_{0,x}\rangle$ and
$|m_{0,y}\rangle$ belong to the first lowest zone:
$m_{0,x}=m_{0,y}=(1,\pm 1)$. Then
${\bf M}^{-1}_{1,xx}={\bf M}^{-1}_{1,yy}={\bf M}^{-1}_1<0$
and the wave is unstable.
The BEC population dynamics in this case is reported in
figure \ref{f7}.
The most interesting feature of the modulational instability developed
is that it evolves in a {\em regular} structure which represents
symmetrically spaced localized in space (we call them
soliton-like) distributions (see figure \ref{f7}b). Each of the
humps shown in the figure represents a tightly confined tube along the
$z$-direction. The number of tubes is proportional to the size
of the box. In order to illustrate the last statement we performed
calculations (see figure \ref{f8}) with parameter settings similar to
those of figure \ref{f7} with the exception of domain size $L=28\pi$.

\begin{figure}[htb]
\centerline{\includegraphics[width=6cm,height=6cm,clip]{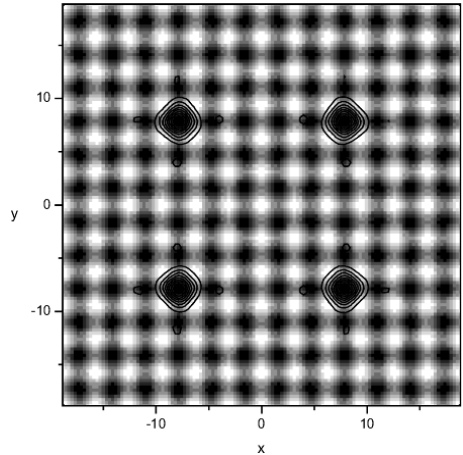}}
\caption{Contour plot of localized excitations
superimposed on the profile of the optical lattice potential, showing
that each excitation fits a single cell. Dark regions correspond to
the minima of the periodic potential for $L_x = L_y = 12\pi$.}
\label{f9}
\end{figure}

In order to understand this behavior we notice that from the
equation (\ref{instab}) we get that the excitations with characteristic
scales $ \lambda>\lambda_{min}=\frac{2\pi}{K_{max}}=
\frac{\pi}{\rho}\sqrt{\frac {|{\bf M}^{-1}_1|}{\tilde{\chi}}}$,
where ${\bf M}^{-1}_1$ is the inverse of the effective mass tensor,
are unstable. The largest increment (i.e. the large Im$|\Omega|$) is
achieved for $\lambda_0=\sqrt{2}\lambda_{min}$. This has two
consequences. First, the symmetry group of the developed structure
must be of ${\bf {\it C}}_n$ type with the symmetry axis
coinciding with that of the condensate, and second, an effective
scale $\lambda_{eff}\sim \lambda_{0}$ must be a characteristic
scale of the most excitation which at the beginning of the evolution.
One can estimate the value of $\lambda_0$ taking into account that for a
chosen point of BZ the inverse of the effective mass tensor is
$|{\bf M}^{-1}_1|\approx 6$ (see figure \ref{f5}) and for the solutions
studied numerically the effective nonlinearity is
$\tilde{\chi}\approx 0.1935$ (the respective normalized eigenfunctions are
approximated by $\frac{2}{\pi}\sin(x)$~\cite{ks}, which gives
$\lambda_0\approx 20.053$. This result corroborates with the distances between the
humps along the radial direction measured from the direct
numerical simulations: $\lambda_{eff}\approx 23.0$ (figure \ref{f9}).
Next we have to take into account that the carrier wave mode is chosen
at the point ${\bf q}=(\pm 1,\pm 1)$ placed at the corner of the BZ which
corresponds to waves whose phases propagate in the directions $x=\pm y$.
This immediately specifies the symmetry ${\bf {\it C}}_4$. In other words,
one can specify the points where the humps
(confined tubes) should appear: in the plane $(x,y)$ these are
intersections of lines $x=\pm y$ with the circles of radii
$\left(\frac 12 +p\right)\lambda_{eff}$ where $p=0,1,...$. In a square box of
size $L$ one will observe $L^2/\lambda_{eff}^2$ humps. This estimate being
rather rough (it does not take into account boundary effects) was
confirmed by our numerical simulations. Also one can  predict that
the characteristic diameters of the humps should be less than
$\lambda_{min}$ ($\approx 7.1$ in our case). This gives an
estimate for the BEC density in a tube $n_t$ versus the initial
density $n_0$: $n_t=\frac{L_xL_y}{\lambda_{min}^2}n_0$, which in
our case gives $n_t\approx 38 n_0$. In order to evaluate the increase
of the BEC atomic density in a soliton-like excitation, we have numerically
integrated $|\psi(x,y,t)|^2$ at $t=85$ (figure \ref{f7}b) over individual
lattice sites. The result is that 65 \% of the BEC matter,
initially uniformly distributed over the optical lattice,
are collected in four sites due to the modulational instability. Therefore,
the increase of the atomic density in a localized excitation is
$n_t = \frac{196}{4} \cdot 0.65 \cdot n_0 \simeq 32 n_0$, wich is close
to the above analytical estimation. For the 3D case considered in the next
subsection this last estimate for the BEC density in hollows $n_h$ is
modified as: $n_h=\frac{L_xL_yL_z}{\lambda_{min}^3}n_0$.

\begin{figure}[htb]
\centerline{\includegraphics[width=8.5cm,height=4.5cm,clip]{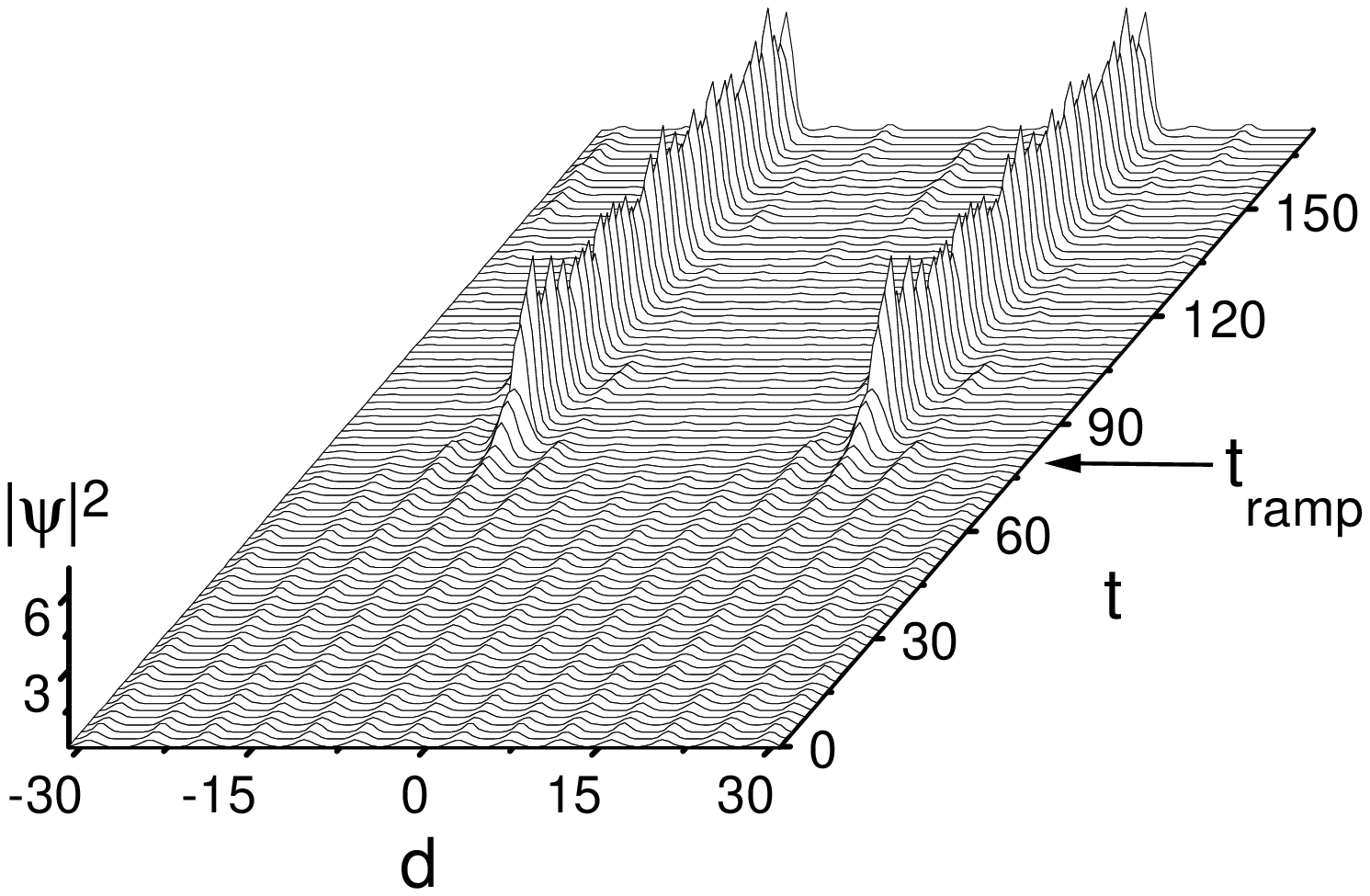}}
\caption{Evolution of the BEC distribution corresponding to
figure \ref{f7}, shown as a diagonal section. $d$ is a distance along the
diagonal of the square domain with $L = 14\pi$.}
\label{f10}
\end{figure}

As we have seen, the modulational instability results in formation of
regular pattern of soliton-like excitations in arrays of
BEC (figures \ref{f7}b, \ 8). However, they eventually decay in accordance
with equation (\ref{NLS}), which does not support stable solitonic solutions
in 2D and 3D. A simple way to retain these excitations would be the
increasing of the strength of the periodic trap potential, when excitations
are formed. High potential barrier between lattice sites then suppresses
the atomic tunneling, providing strong confinement. This idea is illustrated
in figure \ref{f10}, where we show the evolution of the BEC atomic
distribution.
Until $t=85$ the dynamics is guided by the modulational instability,
resulting in formation of regular spatial structures with BEC.
As the soliton-like excitations are formed, the strength of the optical
lattice was increased adiabatically, starting at $t=85$ and ending
at $t=90$, according to law
$A(t)=2 [1-\frac{1}{2}\cos(\frac{\pi(t-t_{\rm ramp})}{\Delta t})]$, where
$t_{\rm ramp} = 85, \quad \Delta t = 5$. To obtain estimates in physical
units recall that time is normalized to $\frac{2m}{\hbar k^2_L}=50 \mu$s
(for $^{87}$Rb atoms and $k_L = 8.06 \cdot 10^{-6}$ m$^{-1}$), and the
strength of the periodic potential is expressed in units of recoil energy
$E_r = \frac{\hbar^2 k_L^2}{2 m}$. Therefore, $t=85$ corresponds to
$\sim$ 4 ms, while $A=1.0$ corresponds to $\sim E_r$.
These values are typical in experimental situations.

To understand the stabilization phenomenon we notice that
$\lambda_{min}$ is of order of $2a_x$ ($2a_y$), which means that
the most of BEC atoms are concentrated in a unique cell
(see figure \ref{f9}). This type of excitations closely resemble
the intrinsic localized modes in BEC in the tight-binding approximation
\cite{trombettoni,abdullaev}. By increasing the potential amplitude one
makes the optical lattice more deep, which in turn leads to
decreasing both the probability of tunneling of atoms from the
most populated cell to neighbor cells and a "number" of atoms in
classically forbidden zone. As a consequence, the BEC density in
the most populated cell is growing, which is illustrated by
figures \ref{f7}c and \ref{f10}.

\subsubsection{Case 2.}

The eigenfunctions ${\Phi}_{m_{0,x}}$ and ${\Phi}_{m_{0,y}}$
belong to different zones, say ${\Phi}_{m_{0,x}}$ belongs to the first
lowest zone: $m_{0,x}=(1,\pm 1)$ and ${\Phi}_{m_{0,y}}$ belongs to the
second lowest zone: $m_{0,y}=(2,\pm 1)$. Then ${\bf M}^{-1}_{1,xx}<0$ and
${\bf M}^{-1}_{1,yy}>0$, and the condensate is unstable.

\begin{figure}[htb]
\centerline{\hspace*{0.5cm}
\includegraphics[width=7.0cm,height=4.0cm,clip]{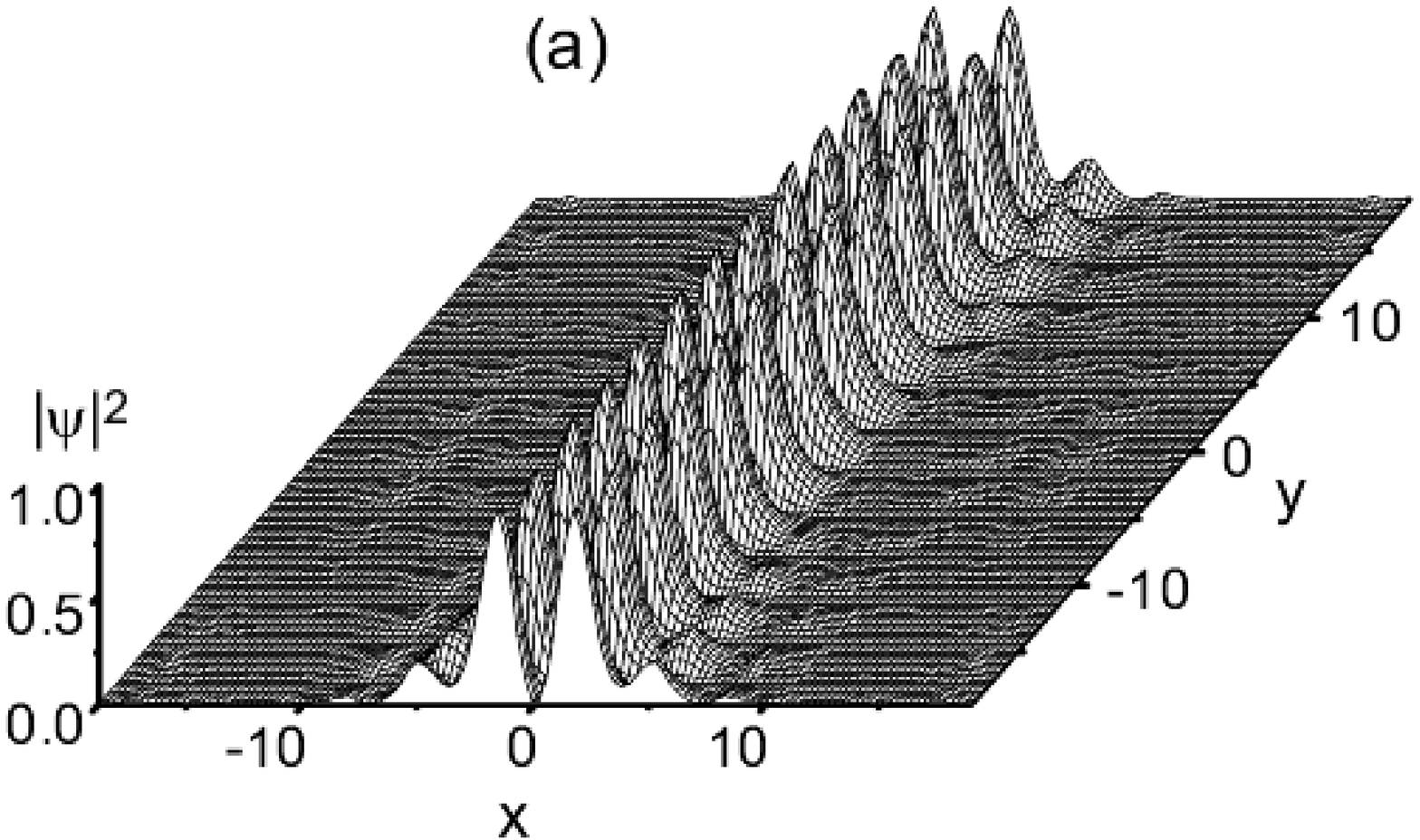} \qquad
\includegraphics[width=6.0cm,height=4.0cm,clip]{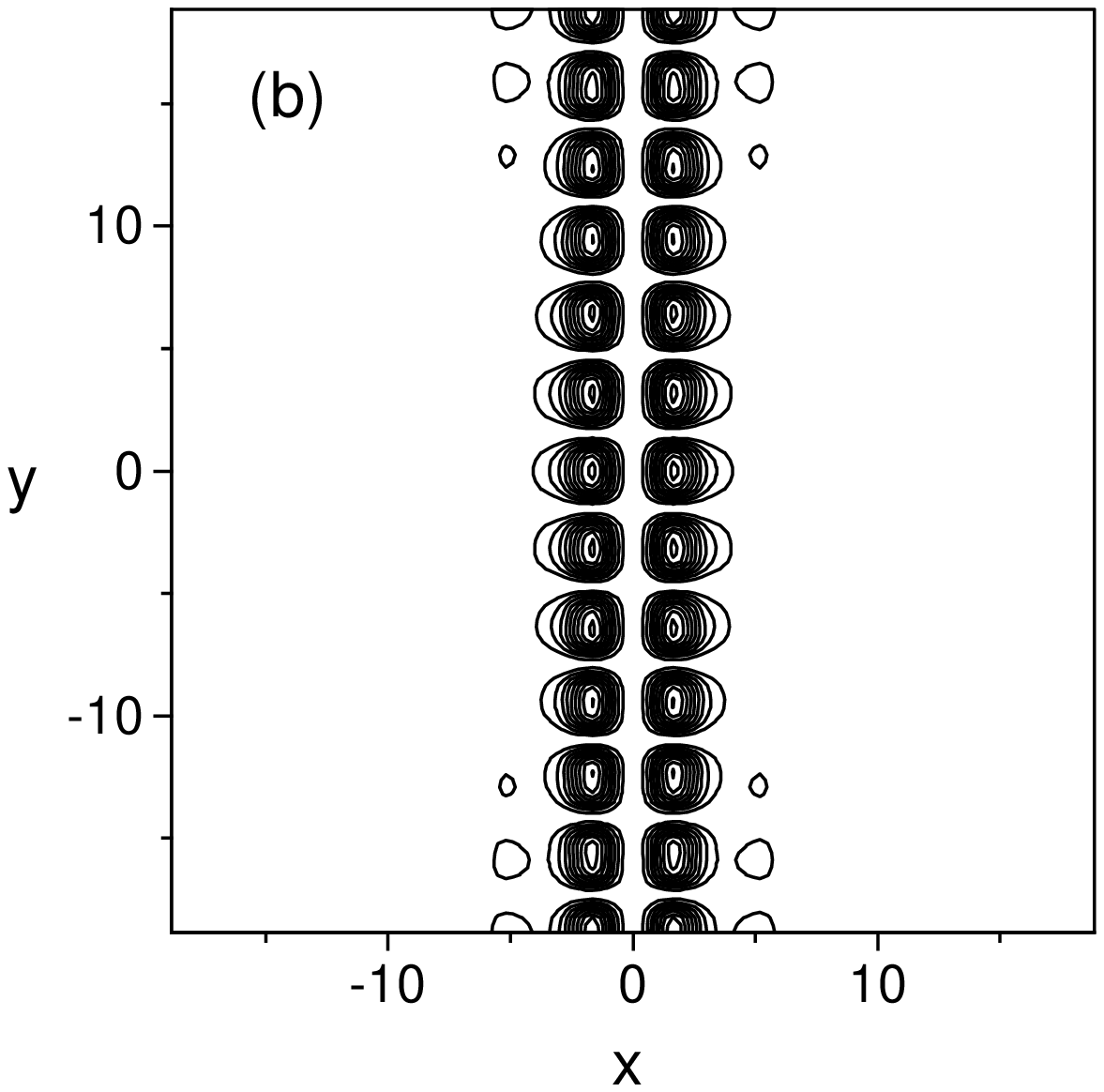}
}
\caption{Distribution of BEC in a 2D optical lattice with
$A=1.0, \ \kappa=2.0, \ \chi=1.0, \ L=12\pi \ $ subject to the initial
condition $\psi(x,y,0) = 0.5 \sin(x) \cos(y)$.
(a) Regular pattern of soliton-like excitations are formed at t =32.
(b) Contour plot of localized excitations showing that they occupy the
sequence of single lattice sites along the $y$ direction. }
\label{f11}
\end{figure}

In this case the instability condition takes the form
$ 0<{\bf M}^{-1}_{2,yy} K_y^2-|{\bf M}^{-1}_{1,xx}| K_x^2 <
4\tilde{\chi}\rho^2 $, and the most unstable excitations have
$K_x^2<\frac{{\bf M}^{-1}_{2,yy}}{|{\bf M}^{-1}_{2,xx} |}K_y^2$
(which is related to the fact that an eigenfunction $\Phi_{m_{0,x}}$
belongs to the "unstable" branch). That is why the main instability
results in a pattern having different symmetry: it develops in the
$x$-direction. Along this direction the pattern is rapidly split
in a sequence of solitary waves. The instability develops also
along $y$-direction, but at much larger time scales (see figure \ref{f11}).
To estimate the number of humps, we take into account that the
periodic boundary conditions impose a characteristic scale
$K_x=2\pi/L$ which leads to the following estimate for the most
unstable scale $\lambda_0$, and thus to $\lambda_{eff}$:
$\lambda_0\approx
2\pi/\sqrt{\frac{2\pi}{L}-\frac{2\tilde{\chi}\rho^2}{{\bf M}^{-1}_1}}$.
For the case, illustrated in figure \ref{f11}a we obtain
$\lambda_0\approx 24$, which yields for the number of humps in
the $x$ direction $L K_{max}/2 \pi \sim L/\lambda_0 \sim 2$.
The instability is developed in
$y$-direction as well, which is characterized by much larger
spatial and temporal scale, and for experimental purposes can be
neglected. In order to preserve the developed spatial structures with
high atomic density, it would be enough to increase the strength of the
periodic trap potential, since the excitations fit the sequence of single
lattice cells along the $y$ direction (figure \ref{f11}b).

\subsubsection{Case 3.}

Both eigenfunctions $\Phi_{m_{0,x}}$ and $\Phi_{m_{0,y}}$
belong to the second lowest zone: $m_{0,x}=m_{0,y}=(2,\pm 1)$.
Then ${\bf M}^{-1}_{2,xx}={\bf M}^{-1}_{2,yy}>0$ and the wave is stable,
which was confirmed numerically using the initial conditions
$\psi(x,y,0)=0.5 \cos(x)\cos(y)$.

\subsubsection{Other points of the Brillouin zone.}

It is of particular interest for applications to explore the development of
modulational instability of initial waveforms, corresponding to different
points of BZ. We have tried the Bloch states with wavevectors spanning the
first BZ. The main observation from the relevant numerical
simulations is that, the modulational instability results in formation of
different spatial structures with BEC depending on the initial waveform.
The time required to formation of these structures is also different.

\begin{figure}[htb]
\centerline{\hspace*{0.5cm}
\includegraphics[width=7.0cm,height=4.0cm,clip]{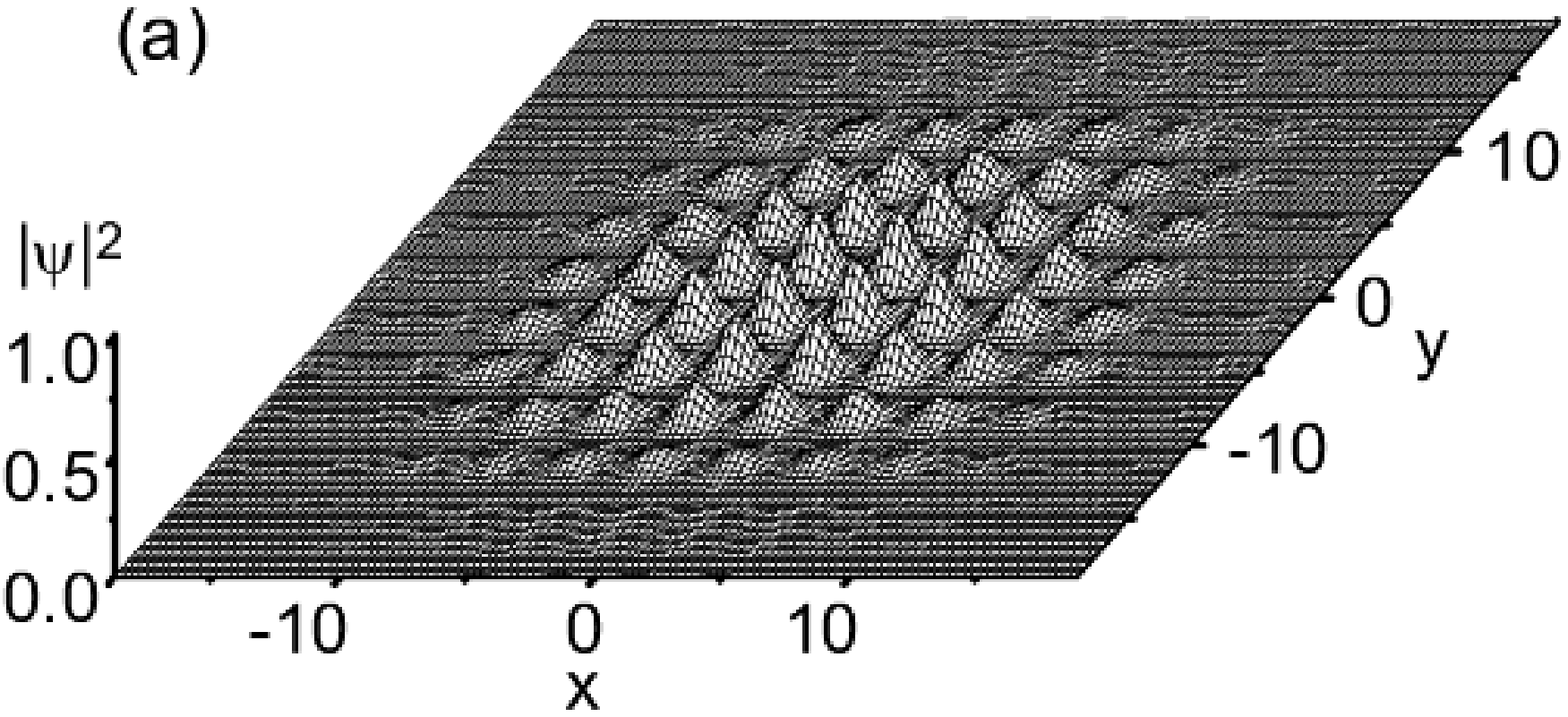} \qquad
\includegraphics[width=7.0cm,height=4.0cm,clip]{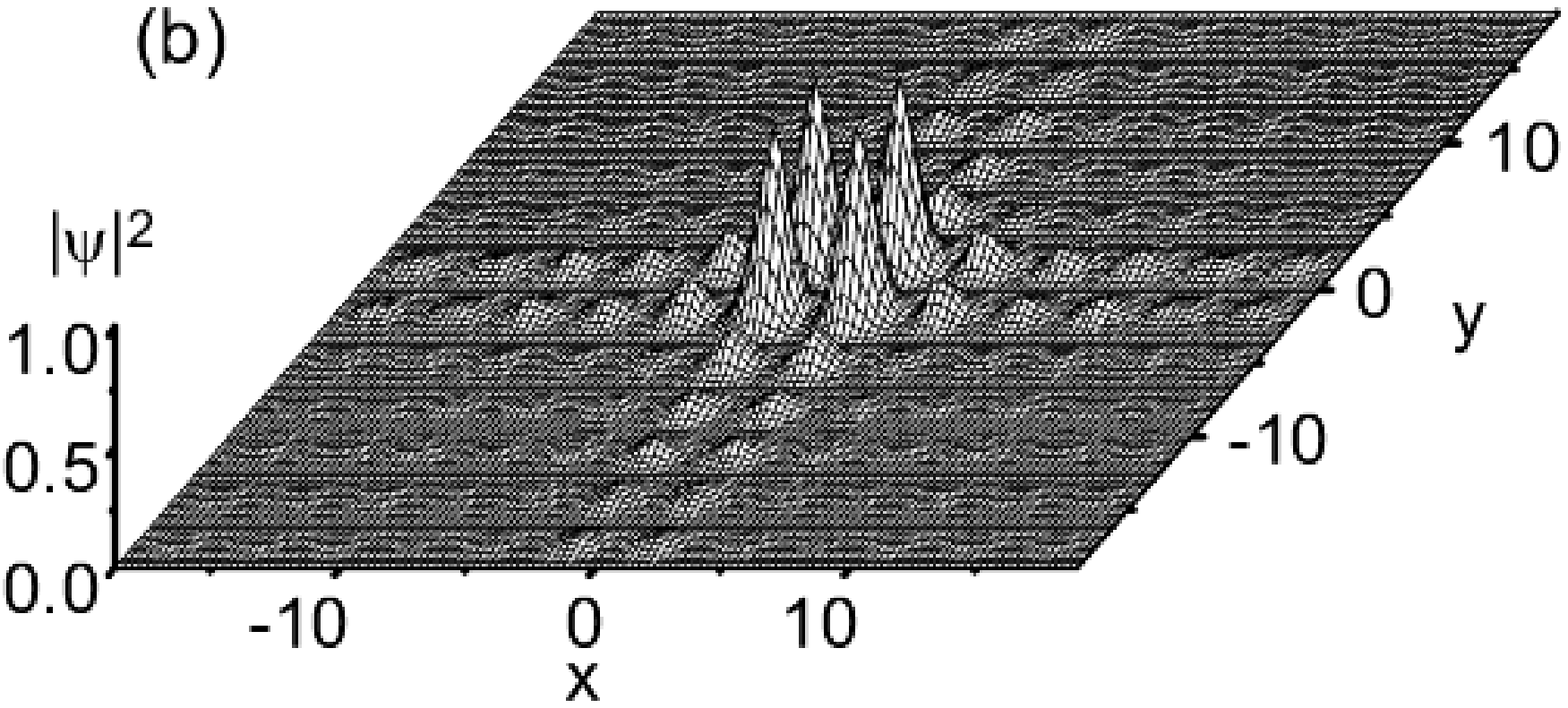}
}
\caption{Redistribution of the BEC atomic population in a 2D optical
lattice due to modulational instability.
$A=1.0, \ \kappa=2.0, \ \chi=1.0, \ L=12\pi \ $
(a) The initial waveform, corresponding to the Bloch state
$m_{0,x}=m_{0,y}=(0.9,\pm 0.9)$.
(b) Formation of soliton-like excitations due to modulational instability
at $t=16$. }
\label{f12}
\end{figure}

As an example, in figure \ref{f12} we report the result of modulational
instability of the Bloch state with $m_{0,x}=m_{0,y}=(0.9,\pm 0.9)$.
This initial distribution seems to be of interest because it
resembles the situation, when a BEC of comparatively small size is
loaded into the optical lattice, so that minor number of lattice
sites are filled in with BEC, others being almost empty. In this
case the cells in the central part of the optical lattice contain
more BEC atoms, than the peripheral ones. The localized
excitations developed due to the modulational instability occupy
the central nearest-neighbor lattice cells.

\subsection{3D optical lattice}
\label{3dcase}

\begin{figure}[htb]
\centerline{\includegraphics[width=6cm,height=6cm,clip]{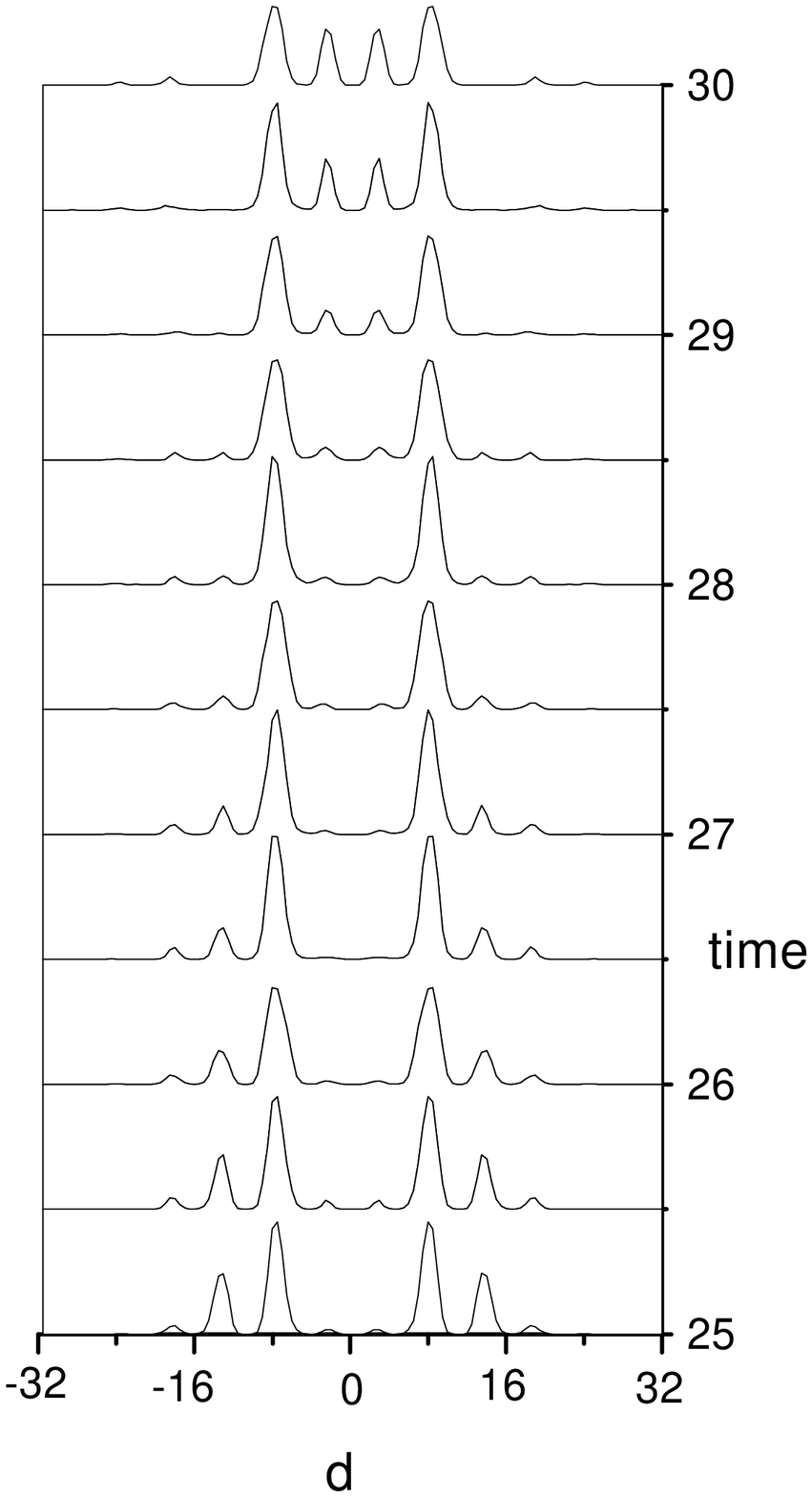}}
\caption{Formation of soliton-like excitations by $t \sim 28$ in a 3D
BEC array. $d$ is the distance along the main diagonal of the cubic domain
with $L = 12 \pi$. The initial condition is
$\psi(x,y,z,0)=0.5 \sin(x) \sin(y) \sin(z)$. }
\label{f13}
\end{figure}

Qualitatively similar behaviour of the modulational instability with respect
to formation of soliton-like excitations was observed in 3D case. The
developed structures are small hollows filled in with BEC atoms of much
greater density compared to surrounding array sites. Figure \ref{f13}
illustrates the emergence of spatial structures with high atomic
concentration in a 3D BEC array, shown as a section along the main
diagonal of the cubic domain with $L = 12 \pi$.
The time interval is selected to display the emergence of soliton-like
excitations at $t \sim 28$, and their subsequent decay.
The relative atomic density in a localized excitation is estimated,
as in the 2D case, by numerical integration of $|\psi(x,y,z,t)|^2$ at $t=28$
over individual lattice sites. In the 3D case $ \sim 67 \%$ of all the
BEC matter, initially uniformly distributed over 1728 lattice sites, are
collected in eight sites due to the modulational instability, therefore
$n_h = \frac{1728}{8} \cdot 0.67 \cdot n_0 \sim 145 n_0$.

For the considered initial parameter settings the long-term evolution
of the atomic distribution over the optical lattice in both 2D and 3D cases
exhibits the recurrence phenomenon, which reveals itself as the return to
the primary low density state. In order to prevent the decay of soliton-like
excitations, similarly to 2D case, the strength of the periodic potential has
to be increased adiabatically when the excitations are formed, e.g.
at $t \sim 28$ for the above 3D parameter settings (figure \ref{f13}).

As far as the initial state is concerned, we remark that it could be
experimentally realized starting from a uniform
condensate with quasi-momentum $k=0$ (i.e. with equally filled
potential wells),  and accelerating the optical lattice in a
particular direction. The acceleration of the lattice, being equivalent
to a gravitational field, will induce the initial $k=0$ state to
move along the band, until it reaches the edge of the band where
the instability develops. Depending on the direction selected in
the real space (this corresponding to a fixed direction in the
reciprocal space), one will get different final localized states
from the band edge instability.
The issues relevant to preparation of particular
Bloch states of the condensate in an optical lattice are discussed in
\cite{denschlag}.

It should be pointed out that the similarity of the development
of modulational instability in all three dimensions is the result of
separability of the periodic trap potential. Although this
corresponds to the majority of experimantal situations, the case of
non-separable potentials (non-orthogonal lattices) represents a significant
interest.

\section{Conclusions}
\label{conclusions}

We have studied the modulational instability in arrays of BEC confined
to optical lattices. The formation of coherent spatial structures with BEC
is shown to be the principal feature of the evolution of atomic distribution
over the optical lattice, when guided by the modulational instability.
In 1D case the developed structures are the matter-wave solitons which
may be regarded as thin disks of highly concentrated BEC atoms. Depending
on significance of the nonlinearity, two distinct types of matter-wave
solitons may develop, these being the envelope solitons (weak nonlinearity)
occupying few lattice sites and intrinsic localized modes
(strong nonlinearity), each fitting a single lattice site. In 2D and 3D
cases the emerging spatial structures with BEC represent soliton-like
excitations regularly arranged over the optical lattice which are, however,
not stable. We proposed a simple way to stabilize these localized excitations
by increasing the strength of the optical lattice when they are formed due to
the modulational instability. Different initial waveforms, corresponding to
particular points on the BZ are tested for instability. The
aspects of developed spatial structures with BEC is shown to depend on the
selected Bloch state. Theoretical model, based on the multiple scale expansion
describes the primary features of emerging soliton-like structures with BEC,
including the number of localized excitations, their spatial simmetry and
relative density of BEC atoms they contain. The proposed method for creation
and preserving of soliton-like spatial structures with highly concentrated
BEC atoms, may be of interest for the physics and applications of BEC.

\ack
V.V.K. acknowledges support from the Programme "Human Potential -
Research Training Networks", contract No. HPRN-CT-2000-00158. M.S.
and B.B. thank the MURST-PRIN-2000 Initiative and the European
grant LOCNET n.o HPRN-CT-1999-00163 for partial financial support.

\end{document}